\documentclass{aa}  
\usepackage{setspace}
\usepackage{array}

\usepackage{verbatim}
\usepackage{graphicx}
\usepackage{txfonts}
\usepackage{xcolor}
\usepackage{xfrac}

\usepackage{ulem} 

\begin{document}

   \title{Long-term scintillation studies of EPTA pulsars.}

   \subtitle{\uppercase\expandafter{\romannumeral1}. Observations and basic results}

   \author{Yulan Liu\inst{1}
          \and
          Joris P. W. Verbiest\inst{1, 2}
          \and
          Robert A. Main\inst{2}
          \and
          Ziwei Wu\inst{1}
          \and
          Krishnakumar Moochickal Ambalappat\inst{1}
          \and
          David J. Champion\inst{2}
          \and 
          Ismaël Cognard\inst{3,4}
          \and
          Lucas Guillemot\inst{3,4}
          \and
          Madhuri Gaikwad\inst{2}
          \and
          Gemma H. Janssen\inst{7,8}
          \and
          Michael Kramer\inst{2}
          \and
          Michael J. Keith\inst{5}
          \and 
          Ramesh Karuppusamy\inst{2}
          \and
          Lars Künkel\inst{1}
          \and
          Kuo Liu\inst{2}
          \and
          James W. McKee\inst{10}
          \and 
          Mitchell B. Mickaliger\inst{5}
          \and
          Ben W. Stappers\inst{5}
          \and
          Golam. M. Shaifullah\inst{9}
          \and 
          Gilles Theureau\inst{3,4,6}
          }

   \institute{Fakultät für Physik, Universität Bielefeld, Postfach 100131, 33501 Bielefeld, Germany
         \and
         Max-Planck-Institut f\"ur Radioastronomie, Auf dem H\"ugel 69, 53121 Bonn, Germany
         \and 
        Station de radioastronomie de Nan{\c c}ay, Observatoire de Paris, CNRS/INSU 18330 Nan{\c c}ay, France
        \and 
        Laboratoire de Physique et Chimie de l'Environnement et de l'Espace, Universit\'e d’Orl\'eans/CNRS, 45071 Orl\'eans Cedex 02, France
        \and
        Jodrell Bank Centre for Astrophysics, Department of Physics and Astronomy, University of Manchester, Manchester M13 9PL, UK
        \and 
        LUTH, Observatoire de Paris, PSL Research University, CNRS, Universit\'e Paris Diderot, Sorbonne Paris Cit\'e, 92195 Meudon, France
        \and
        ASTRON, Netherlands Institute for Radio Astronomy, Oude Hoogeveensedijk 4, 7991 PD, Dwingeloo, The Netherlands
        \and 
        Department of Astrophysics/IMAPP, Radboud University Nijmegen, P.O. Box 9010, 6500 GL Nijmegen, The Netherlands
        \and 
        Dipartimento di Fisica ``G. Occhialini'', Universit\`a di Milano-Bicocca, Piazza della Scienza 3, 20126 Milano, Italy
        \and 
        Canadian Institute for Theoretical Astrophysics, University of Toronto, 60 Saint George Street, Toronto, ON M5S 3H8, Canada
        }

   \date{}

 
  \abstract
   {Interstellar scintillation analysis of pulsars allows us to probe the small-scale distribution and inhomogeneities of the ionized interstellar medium. From the frequency scale of scintillation, one can estimate the geometric time delays from multipath propagation, a source of (typically) unmodeled, correlated noise in pulsar timing. Interstellar scintillation analysis of well-timed pulsars is useful to quantify the effects of time delays and may lead to improved timing precision, enhancing the probability of detecting gravitational waves.
   }
  {Our priority is to present the data set and the basic measurements of scintillation parameters of pulsars, employing long-term scintillation observations carried out from 2011 January to 2020 August by the European Pulsar Timing Array radio telescopes in the 21-cm and 11-cm bands. Additionally, we aim to identify future possible lines of study using this long-term scintillation data set.
  }
  {The autocorrelation function of dynamic spectra has been used to estimate the scintillation bandwidth $\nu_{\rm d}$ and scintillation timescale $\tau_{\rm d}$.}
   {We present the long-term time series of $\nu_{\rm d}$ and $\tau_{\rm d}$ for 13 pulsars. Sanity checks and comparisons indicate that the scintillation parameters of our work and previously published works are mostly consistent. For two pulsars, PSRs~J1857+0943 and J1939+2134, we were able to obtain measurements of the $\nu_{\rm d}$ at both bands, which allowed us to derive the time series of frequency scaling indices with a mean and a standard deviation of 2.82$\pm$1.95 and 3.18$\pm$0.60, respectively. We found some interesting features which will be studied in more detail in subsequent papers in this series: (i) in the time series of PSR~J1939+2134, where $\nu_{\rm d}$ and $\tau_{\rm d}$ sharply  decrease associated with a sharp increase in the dispersion measure; (ii) PSR~J0613$-$0200 and PSR~J0636+5126 show a strong annual variation in the time series of the $\tau_{\rm d}$; and (iii) PSR~J1939+2134 shows a weak anticorrelation between the scintillation timescale and the dispersion in Westerbork Synthesis Radio Telescope data.
   }
   {}

   \keywords{Pulsar: general --
                ISM: general}
\maketitle
%

\section{Introduction}
\label{sec:introduction}
Radio pulsars, rapidly rotating and highly magnetized neutron stars, have been a fascinating subject of research since their discovery in 1967 \citep{hbp+68}. As very compact sources of radio emission, they are also good probes of the interstellar medium (ISM), in particular its ionized component. 
There are four main propagation effects that occur when the signals from pulsars pass through the ionized interstellar medium (IISM): frequency dispersion, Faraday rotation, interstellar scintillation (ISS), and pulse broadening (see \citealt{lk+12}, for a review). 
Frequency dispersion and Faraday rotation effects can both be understood by propagation through a homogeneous medium. However, the IISM in general is inhomogeneous and highly turbulent \citep{ric90}, resulting in ISS or pulse broadening \citep{sch68}. 
Phase differences in the deflected pulsar signals result in interference observed as intensity fluctuations in frequency and time, similar to the familiar optical "twinkling" of stars caused by the atmosphere of the Earth.  

The density inhomogeneities of the IISM at various length scales leads to two different regimes of scintillation, referred to as "weak" and "strong" scintillation.
Weak scintillation is weakly modulated, implying that most of the light has traveled almost a single path and that phase perturbations at the observer's plane are small. Strong scintillation is fully modulated, implying that most of the light arrives via multiple paths and that phase perturbations at the observer's plane are large. \citet{ric90} provided a scintillation modulation index $m$ that is the standard deviation of the observed flux densities divided by their mean. In weak scintillation, $m$ is much less than unity; in strong scintillation, $m$ is approximately unity or larger than unity.
The variations of strong scintillation emerge on two different timescales: the short-term variability (on the order of minutes) caused by changing interference between parts of the scattering disk and long-term variability (on the order of months) with spatial scales on the order of the scattering disk so that observers receive light from a fully different set of scattering points. In this paper, we investigate the short-term variability of the strong scintillation exploiting the long-term observations of pulsars from the European Pulsar Timing Array (EPTA).

Pulsar scintillation properties are best studied using their dynamic spectrum, which is the two-dimensional image of the pulsed intensity as a function of observing time and observing frequency. The interference maxima in the dynamic spectrum are called scintles, where broad and narrow scintles indicate small and large separations of the ray paths, respectively.
There are two common approaches that are described in the following to further study scintillation properties. Firstly, employing two parameters, the scintillation bandwidth $\nu_{\rm d}$ and the scintillation timescale $\tau_{\rm d}$, the average characteristics of scintles can be quantified for each observation. 
$\tau_{\rm d}$ is the half width at 1/e along the time axis and $\nu_{\rm d}$ is the half width at half maximum along the frequency axis in the two-dimensional autocorrelation function (ACF) of the dynamic spectrum \citep{cor86}.
Secondly, a "criss-cross" pattern can be seen in the dynamic spectrum of some observations \citep{hew80}. In order to investigate such "slopes" or "criss-cross" scintles, \citet{cw84} presented a secondary spectrum by taking the two-dimensional Fourier transform of the dynamic spectrum.

\citet{smc+01} found a faint but clear arc in the secondary spectra of several pulsars at 430\,MHz, using high-resolution, high-sensitivity dynamic spectra. 
\citet{wmsz04,crs+06,pl+14,gs+19} elaborated the theory of parabolic arcs in the secondary spectrum. 
\citet{rcb+20}, \citet{msa+20} and \citet{mma+22} derived the location and the nature of the scattering screen from fitting arc curvatures for PSRs~J0437$-$4715, J0613$-$0200 and J1643$-$1224, respectively.
\citet{yzm+21} reported a evidence for three-dimensional alignment between the spin and velocity vectors based on scintillation arc curvature studies. 
However, the majority of observations in this work do not show meaningful scintillation arcs. We restrict the analysis to the $\nu_{\rm d}$ and the $\tau_{\rm d}$, which are easier to measure.

In general, $\nu_{\rm d}$ is highly frequency dependent. Thus, gathering measurements at multiple frequencies allows one to estimate the scaling index $\alpha$ of the scintillation bandwidth with frequency. By establishing the relation between the scaling index and the spectral index $\beta$ of the electron density fluctuation spectrum versus wavenumber, the turbulence characteristics of the IISM can be inferred \citep{gn85}. 
\citet{ric77} provided that the scaling index of the scintillation bandwidth with frequency is $\alpha$ = 4.0 in the thin screen model.
Another more commonly used scaling index is $\alpha$ = 4.4 for Kolmogorov turbulence with a spectral index $\beta$ = 11/3 in a uniform medium \citep{rnb86}. \citet{gup00} summarized which scintillation studies favoured, or disfavored, a pure Kolmogorov spectrum. To explain events that disfavored a pure Kolmogorov spectrum, \citet{lr99,lr00} presented a steeper spectrum with a spectral index of $\beta=4$, this type of spectrum could be caused by a medium with abrupt density changes.

However, $\tau_{\rm d}$ is weakly frequency dependent but more modulated by the transverse velocity of the pulsar, Earth and IISM. \citet{ls82} presented some observations which show that the scintillation characteristics of pulsars are closely related to their velocities, as measured by their proper motions.
In turn, measurements of scintillation also can be used to estimate the pulsar's transverse velocity. Subsequently, 
\citet{gup95} presented a correlation between pulsar proper motion velocities and scintillation velocities of radio pulsars, and \citet{cr98} extended it for different scattering geometry models.
In some cases, the time series of $\tau_{\rm d}$ show an annual variation, and for binaries, it is possible to show orbital-annual variations. By employing these variations, the small-scale distribution and inhomogeneities of the IISM, or some orbital parameters can be determined. 
For examples, \citet{lrr+08} investigated the transverse velocity and the distribution along the line of sight of plasma clouds using annual variations from three quasar radio sources,
\citet{rcn+14} exploited annual and orbital variations of scintillation timescales in the double pulsar system J0737$-$3039 to estimate the location, spatial structure and transverse phase gradient of a scattering screen as well as two orbital parameters: the inclination angle $i$ and the position angle of the ascending node $\Omega$. 
\citet{rch+19} also used the annual and the orbit variations to resolve the previous ambiguity in the sense of the $i$, the $\Omega$ and measurements of the spatial structure of the interstellar electron-density fluctuations, the distribution of scattering material along the line of sight, and spatial variation in the strength of turbulence from epoch to epoch.

We describe our observations and analysis methods in Section~\ref{sec:data} and Section~\ref{sec:methods}, respectively. The measurements of scintillation parameters are presented in Section~\ref{sec:results} and the comparison between our results and previous studies and NE2001 are given in Section~\ref{sec:comparison}. In Section~\ref{sec:discussion}, we discuss the significance of our results in the context of Pulsar Timing Arrays (PTAs). We present the conclusion and further research in the last section.

\section{Data}
\label{sec:data}    
The EPTA \citep{dcl+16} is a collaboration of European research institutes and radio observatories that was established in 2006 and makes use of the five largest telescopes at decimetric wavelengths in Europe:
the Effelsberg 100-m Radio Telescope (EFF), the Lovell Radio Telescope at the Jodrell Bank Observatory (JBO), the Nan{\c c}ay radio telescope (NRT), the Westerbork Synthesis Radio Telescope (WSRT) and the Sardinia Radio Telescope (SRT). 
The EPTA aims to study the astrophysics of millisecond pulsars and to detect cosmological gravitational waves in the nanohertz regime. 

In this work, we aim to analyze long-term scintillation observations for EPTA pulsars. To define the sample of pulsars, we considered all EPTA sources included in \citet{dcl+16} that were expected to have an easily resolvable scintillation bandwidth, either based on earlier scintillation studies, or based on model predictions given by the NE2001 model \citep{cl02}. This resulted in the 13 MSPs listed in Table~\ref{tab:scinpara}. 
They span spin periods from 1.56 to 16.45\,milliseconds with a DM range of $9-71$\,pc\,cm$^{-3}$.
The observations used for those 13 pulsars are mainly from continued monitoring of the NRT telescope that has the largest observing bandwidth, and partly from the EFF, JBO and WSRT telescopes. As the SRT has only recently commenced MSP monitoring, no data from this telescope are included. We describe the observations used in this paper from each telescope in the following subsections, and the relevant details of the observations after removal of the radio-frequency interference (RFI) are included in Table~\ref{tab:scinpara}.

\subsection{Nan{\c c}ay radio telescope}
The 94-m equivalent NRT is located in the small commune of Nan{\c c}ay, two hours' drive south of Paris, France. Two receivers: the low-frequency receiver (1.1\,-\,1.8\,GHz)  and the high-frequency receiver (1.7\,–\,3.5\,GHz) with a total observable bandwidth of 512\,MHz centered on a selectable center frequency are used, most of the observations with the low-frequency receiver were carried out at a center frequency of 1484\,MHz. 
All NRT data in this work was recorded with the new wide-band NUPPI dedispersion backend \citep{ldc+14} that has been using since August 2011.
Data are recorded with a subintegration length of 15, 31 or 61 seconds and a channel bandwidth of 4\,MHz. The observations used in this work were made between 2011 August and 2020 January. Each observation's duration is around one hour.

\subsection{Effelsberg Radio Telescope}
The EFF 100-m radio telescope of the Max-Planck Institute for Radioastronomy (MPIfR) is located about 1.3\,km northeast of Effelsberg, a southeastern part of the town of Bad Münstereifel in Germany. Observations used in this paper come from two receivers: a single-pixel receiver with a center frequency of 1347.5\,MHz and 200\,MHz of bandwidth (observations prior to 2017), and a multibeam receiver with a center frequency of 1397.5\,MHz and a bandwidth of 400\,MHz (2017 onward). 
Data are recorded with a channel bandwidth of 1.5625\,MHz and a subintegration length of 10\,seconds. The data used in this work have been regularly collected by the PSRIX (200-MHz data) and Automatix (400-MHz data) systems \citep{lkg+16} from 2011 March to 2020 June with typical observation lengths of 20-50 minutes. As scintillation strength is highly frequency-dependent and the data were not averaged in frequency, in order to produce as homogeneous a data set as possible, all observations observed with the multibeam receiver were downsized to 200\,MHz of bandwidth with a center frequency of 1347.5\,MHz by cutting out the edges of the band.
Additionally, some pulsars are observed at the 11-cm band with a center frequency of 2627\,MHz and a bandwidth of 200\,MHz from 2011 March to 2015 July, data for those pulsars are recorded with a channel bandwidth of 1.5625\,MHz and a subintegration length of 10\,s.

\subsection{Lovell Radio Telescope}
The Lovell Telescope is a 76.2-m radio telescope at the JBO, near Goostrey, Cheshire in North West  England. Pulsars in this work are regularly observed and we used the data from 2011 April to 2020 January.
All observations are recorded with a bandwidth of 400\,MHz, a channel bandwidth of 1.5625\,MHz and a subintegration time ranging from 10 to 120 seconds. Observation durations are approximately 30 minutes.

\subsection{Westerbork Synthesis Radio Telescope}
The WSRT is an aperture synthesis interferometer located north of the village of Westerbork in the northeastern Netherlands. It consists of a 2.7-km east-west array of fourteen 25-m dishes, adding up to a collecting area equivalent to that of a 94-m dish when combined as a tied array. We only use data on three pulsars from this telescope, one (PSR~J0636+5128) was observed from 2014 January until 2015 June with an observation length of one hour, a subintegration time of 60\,seconds, a bandwidth of 80\,MHz centered on a frequency of 350\,MHz and a channel bandwidth of 0.156MHz. The other two pulsars (PSRs~J1939+2134 and J1713+0747) were observed from 2007 February until 2015 June for around 30\,minutes per observation, with the same subintegration time of 60\,seconds, and a bandwidth of 160\,MHz centered on a frequency of 1380\,MHz and with a channel bandwidth of 0.3125\,MHz.


\section{Methods}\label{sec:methods}
\begin{figure*}
\centering
\includegraphics[width=9cm]{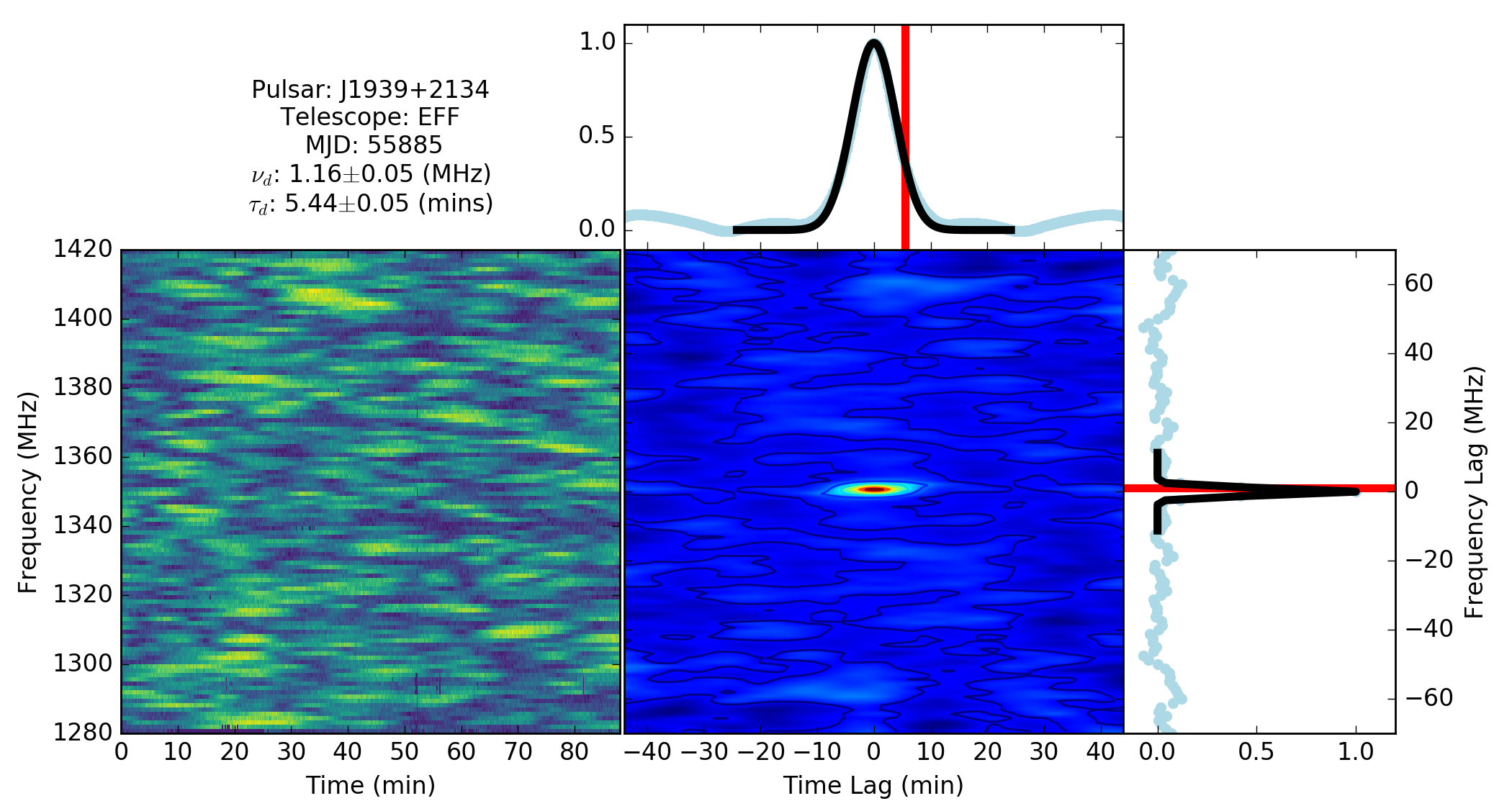}
\includegraphics[width=9cm]{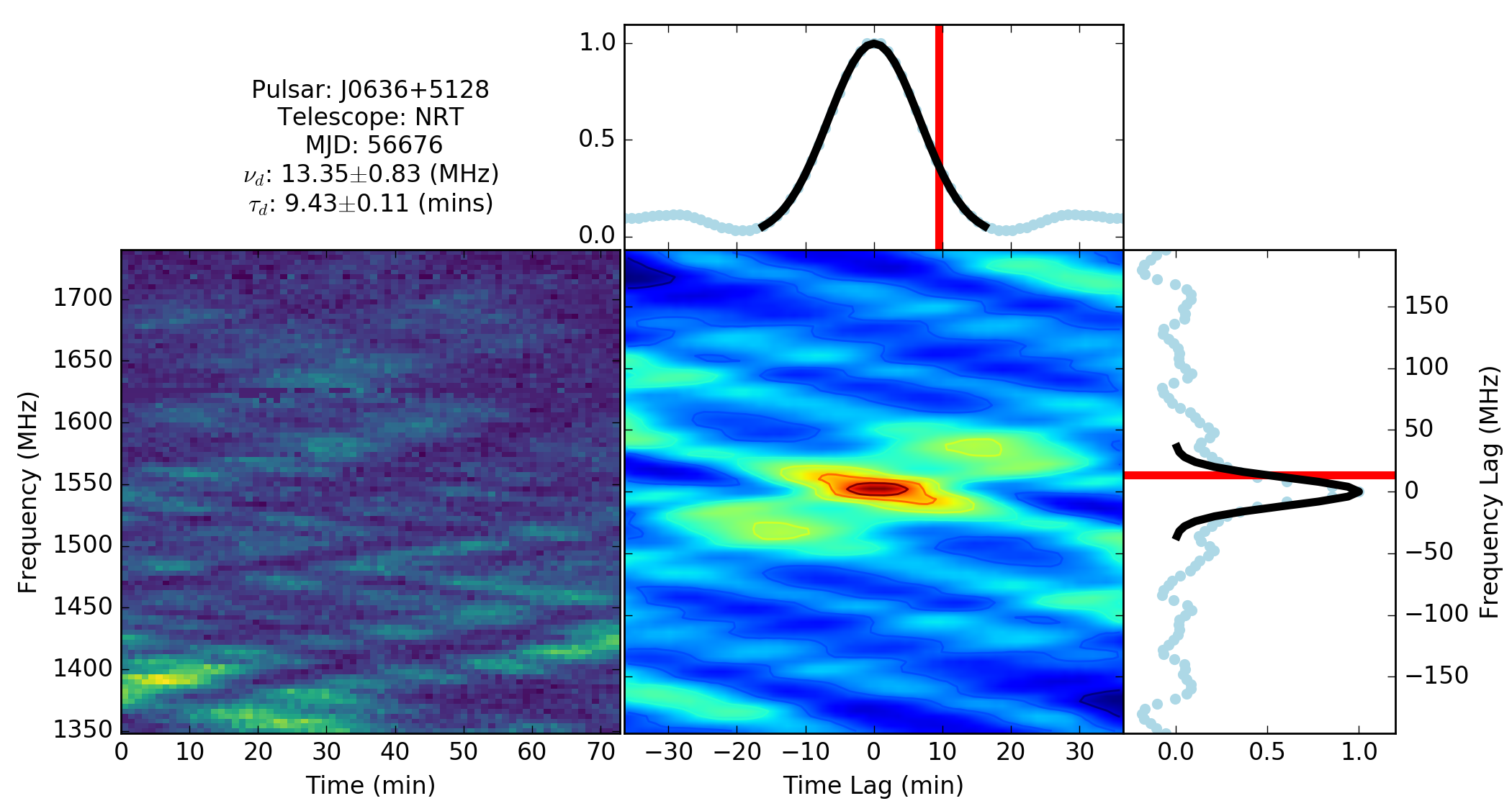}
\includegraphics[width=9cm]{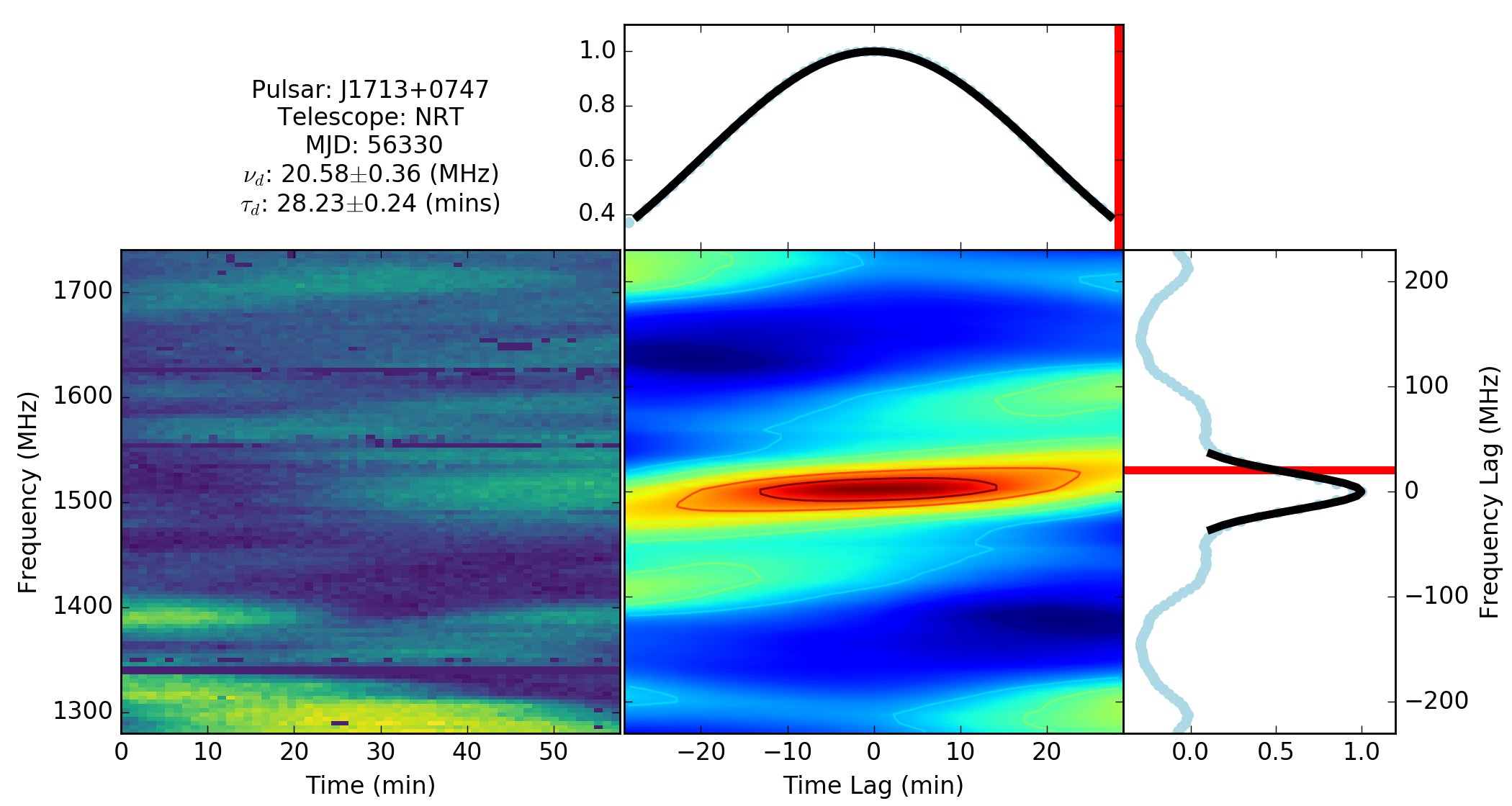}
\includegraphics[width=9cm]{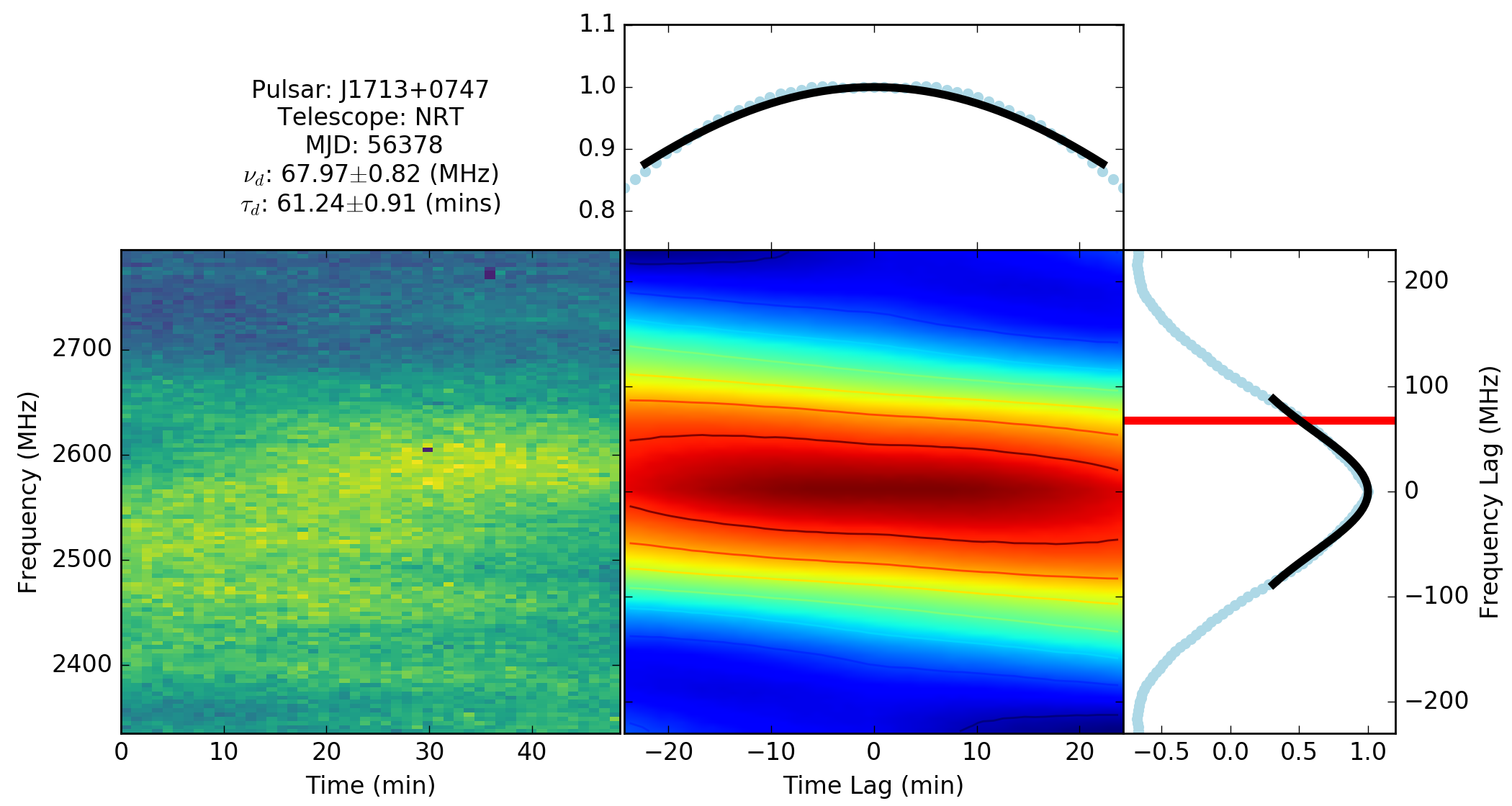}
\caption{Dynamic spectra $I (f,t)$ from four observations with different scintle sizes in the left panels and their normalized autocorrelation functions in the right panels. The horizontal and vertical axes correspond to the time (or time lag) and frequency (or frequency lag), respectively. In the two smaller side plots, the sky blue points are the 1D autocorrelation functions at zero time and frequency lag, the black lines are the best Gaussian fits, the red lines indicate the values for $\nu_{\rm d}$ and $\tau_{\rm d}$. The pulsar name, telescope, epoch of observation, $\nu_{\rm d}$ and $\tau_{\rm d}$ are given above the dynamic spectra.}
      \label{fig:dyacf}
   \end{figure*}
\subsection{Scintillation parameters}

If radio-frequency interference (RFI) is impulsive and broadband or narrow-band and persistent, it will show up as an extended feature in the dynamic spectrum -- in which case it is likely to have a strong influence on the derived scintillation parameters. 
Therefore, before getting the dynamic spectrum,  RFI rejection is necessary. To distinguish RFI from pulsar signals in each time and frequency bin, we calculated the pulse profile's Pearson autocorrelation coefficient in the off-pulse area with a phase lag of one bin.
The pixels in the dynamic spectrum with an autocorrelation coefficient at least $L \sigma$ above the mean (where $\sigma$ is the standard deviation of the correlation coefficients obtained), were identified as RFI and replaced with zeroes. $L$ is a tunable parameter in the range $\sim 1.1-2$, which was manually adjusted for optimal performance as it is strongly dependent on the amount of RFI present. In addition the highest and lowest frequency channels were removed since a combination of decreasing telescope sensitivity and increased potential of systematic effects made these channels unreliable. In consequence, the effective center frequency and bandwidths given in Table~\ref{tab:scinpara} differ from those of the raw data described in Section~\ref{sec:data}.

After dedispersing, we formed total intensity (Stokes I) profiles for each channel and subintegration by summing polarizations, producing a data cube $I\,(f, t, phase)$ from each folded archive.
A template was obtained by averaging all intensity profiles over the entire observing duration and over the usable range of the observing bandwidth for every observation. 
The on- and off-pulse regions were determined by the signal-to-noise ratios (S/Ns) of the template bins, with the on-pulse region having ${\rm S/N}>7$, and vice-versa for the off-pulse region with ${\rm S/N} \le 7$.

To acquire only scintillation patches in the dynamic spectrum, we removed the frequency dependence of our observational sensitivity, which is caused by the combination of instrumental sensitivity and pulsar spectral index. 
To remove the instrumental sensitivity effects in different frequency channels, we divided the intensity in each channel by the time average of the off-pulse region across the full observation. 
However, some bright pulsars have emission in the off-pulse region, which contributes to off-pulse emission getting confused with scintillation \citep{rd+18}. 
Thus, we averaged many observations to remove the instrumental bandpass for the bright pulsars. 
After removing the instrumental bandpass, we averaged the power of the on-pulse region based on as many observations as possible to calculate and then remove the effect of the pulsar's spectral index. Furthermore, in each time and frequency bin, we subtracted the mean of the off-pulse region to remove variable background flux.
Finally, the remaining power in the on-pulse region in each time and frequency bin was used as a point in the dynamic spectrum. 
Four examples of normalized dynamic spectra are presented in the left panels of Figure~\ref{fig:dyacf}, showing the wide range of scintle sizes between our sources.

Following \citet{cor86}, in order to quantify the average characteristics of scintles at each observation, $\tau_{\rm d}$ and $\nu_{\rm d}$ have been defined, where $\tau_{\rm d}$ is the half width at 1/e along the time axis and $\nu_{\rm d}$ is the half-width at half maximum along the frequency axis in the two-dimensional ACF of the dynamic spectrum. \citet{grl94} defined the ACF as
\begin{equation}
\label{eq:acf}
{\rm ACF}(\Delta{f},\Delta{t}) = \frac{R(\Delta{f},\Delta{t})}{R(0,0)},
\end{equation}
where $R(\Delta{f},\Delta{t})=<\triangle I(f,t) \cdot \triangle I(f+\Delta{f},t+\Delta{t})>$, $\triangle I(f,t)$ is the deviation of the intensity from the mean intensity < $I$ >  of the dynamic spectrum, and < > denotes an average over the range in frequency and time from one observation. In the calculation of the global mean intensity of the dynamic spectrum  < $I$ >, we masked all zeros that resulted from RFI removal. 
Then, we used a convolution algorithm that utilises the fast Fourier transform to do the calculation of the ACF.
Due to the finiteness of the data, the mentioned convolution algorithm automatically padded the arrays $\triangle I(f,t)$ and $\triangle I(f+\Delta{f},t+\Delta{t})$ with zeros, which results in a different number of samples contributing to various pixels of the ACF. To correct this, we normalized the ACF to take into account the different numbers of contributions to each pixel.
Some pulsars, especially weaker sources, show a sharp spike at zero frequency and time lag of the ACF, resulting from the noise in the dynamic spectrum.
To remove that spike, we replaced it with the mean value of the nearest two points along the time lag axis.

As in previous works, cuts along two axes of the ACF (called 1D ACF) and a  Gaussian function \citep{grl94} were used to obtain the $\nu_{\rm d}$ and $\tau_{\rm d}$,
\begin{equation}
\begin{split}
\label{eq:acfshape}
& {\rm ACF}(\nu=0,\tau)={\rm exp}(-a \cdot \tau^2),\\
& {\rm ACF}(\nu,\tau=0)={\rm exp}(-b \cdot \nu^2).
\end{split}
\end{equation}

We determined the fitting range from the center of the ACF to the point where the uncertainty of the fitting parameters from the least-squares fits is the smallest.
For the determination of scintillation timescales, we tried the theoretical form for a Kolmogorov scattering medium ${\rm ACF}(\nu=0,\tau)= {\rm exp}(-a \cdot \tau^{5/3})$ \citep{cmr+05,crg+10,rch+19}. However, we found that the commonly used Gaussian function gives a better fit to the shape along the time axis of the ACF of our pulsars. 

Then, the scintillation parameters are derived with
\begin{equation}
\begin{split}
\label{eq:scinpara}
&\tau_{\rm d}=\sqrt{\frac{1}{a}}, \\
&\nu_{\rm d}=\sqrt{\frac{\ln{2}}{b}}.
\end{split}
\end{equation} 

Usually, the uncertainty of the scintillation parameters consists of the formal uncertainty from the fitting procedure and the statistical uncertainty due to the finite number of scintles in the dynamic spectrum. 
Following \citet{cwb85}, the statistical fractional uncertainty, $\sigma_{\rm est}$, is given by
\begin{equation}
\label{eq:error_sys}
\sigma_{\rm est}=\left[f_{\rm d}\cdot \left(\frac{\rm{BW} \cdot \rm{t_{obs}}}{\nu_{\rm d} \tau_{\rm d}}\right)\right]^{-0.5},
\end{equation}
where BW is the bandwidth of the observation, $\rm{t_{obs}}$ is the length of the observation and $f_{\rm d}$ is the filling factor. The filling factor is an estimate of the fraction of the dynamic spectrum that is filled with scintles. For our 13 pulsars, depending to the size of their scintles, we used filling factors from 0.2 to 0.4 for different pulsars, please see the details in Table~\ref{tab:scinpara}. We plotted the ACF of four representative dynamic spectra in the middle panel of Figure~\ref{fig:dyacf}. In the side panels, cuts along the two axes of the ACF are plotted, the black lines are the best Gaussian fits.

However, as can be seen in Figure~\ref{fig:dyacf}, since the samples in the ACF are highly correlated, the uncertainties derived from the ACF are heavily underestimated. In order to reflect the true Gaussian variations (as a proxy for the measurement uncertainty) of measurements in Figure~\ref{fig:TS}, we
\begin{enumerate}
    \item took the Fourier Transform (FT) and then the power spectrum (PS) of the time series of the $\nu_{\rm d}$ (or $\tau_{\rm d}$);
    \item used the PS to identify frequencies with excess power;
    \item removed these frequency channels from the FT, in practice, the remaining FT now describes the white noise (WN) in our data without containing any of the non-white variations; 
    \item determined the inverse Fourier Transform of the WN back to the time domain;
    \item calculated the Root Mean Square (RMS) of the whitened data;
    \item added that RMS in quadrature to the error bars.
\end{enumerate}
In all cases this additional measurement uncertainty dominated the resulting error bar of the measurements.

\subsection{The time series of DM}
Fluctuations in electron density that give rise to scattering should also contribute to variations in DM.  If the scattering angle is small, as is typical in the 21-cm band, changes in DM and scintillation are likely to happen at roughly the same time. 
We therefore expected to find some relations between scintillation parameters and DM.
For example, \citet{mls+18} demonstrated a correlation between scattering and dispersion measure variations in observations of the Crab pulsar (with a correlation coefficient of 0.56 ± 0.01), and \citet{cks+15} reported two interesting events showing increases in both the intensity of scintillation and dispersion toward PSRs~J1603$-$7202 and J1017$-$7156, which they referred to as extreme scattering events (ESEs).

Thus, the observations with RFI removed are used to get the time series of DM by using the \textsc{PSRCHIVE} \citep{hsm+04} and \textsc{TEMPO2} \citep{hem06} software packages. We first used a fully frequency-averaged high-S/N observation to produce an analytic template by employing the \textsc{PSRCHIVE} program \textsc{paas}. Dependent on the mean S/N of each pulsar, then, all observations are scrunched to 10-20 channels. The analytic template and the \textsc{PSRCHIVE} program \textsc{pat} are used to yield ToAs, whereby each channel yields one ToA. With the \textsc{TEMPO2} software package and a good-quality timing model \citep[taken from][]{dcl+16}, for each pulsar, we removed all ToA outliers and all ToAs with large error bar (several times larger than the median). Finally, the remaining ToAs of each observation are used to fit for the DM at that observing epoch.

The differential time delay across the effective bandwidth, which is naturally related to the DM measurement,
\begin{equation}
\label{eqn:dm} 
\Delta{t} \propto {\rm DM} \frac{2 {\rm BW} }{f_{\rm c}^3},
\end{equation}
\noindent where BW and $f_{\rm c}$ are the bandwidth and center frequency in\,MHz,  $\Delta{t}$ is expressed in seconds and DM is in units of pc\,cm$^{-3}$ \citep{vs+18}. That means the wider BW and the lower center frequency can yield the higher precision DM measurement.
For our EPTA data, therefore, a wide range in DM precision can be obtained by fitting DM to individual observations. Consequently, we only use the most precise DM measurements obtained, which are typically from 21-cm Nan{\c c}ay data.


\section{Results}\label{sec:results}
\begin{table*}\tiny
\centering
\caption{Measured scintillation parameters and observation information}
\label{tab:scinpara} 
\renewcommand\arraystretch{1.5}
\begin{tabular}{cccccccccccccc}
	\hline 
	\hline
Pulsar &$f_{\rm d}$ & Telescope & $f_{\rm c}$ & MJD range & N$_{\rm{obs}}$ & CHBW & BW & t$_{\rm{sub}}$ & $\bar{\rm{t}}_{\rm{obs}}$ & $\nu_{d}$&$\tau_{d}$&$\tau_{st}$\\
	&&&{\small(MHz)}&&&{\small(MHz)}&{\small(MHz)}&{\small(sec)}&{\small(min)}&{\small(MHz)}&{\small(min)}&{\small(ns)}\\
\hline
J0023+0923&0.3&NRT&1484&55857--58125&61&4&512.0&61&55&$47^{88}_{24}$&$\textcolor{red}{57}^{\textcolor{red}{101}}_{32}$&$3.9^{7.8}_{2.1}$\\
\\
J0613-0200&0.2&EFF&1349&55661--58019&55&1.56&140.6&10&34&$\textcolor{red}{1.7}^{3.3}_{\textcolor{red}{1.1}}$&$11^{24}_{7}$&$\textcolor{red}{110}^{\textcolor{red}{170}}_{60}$\\
$\cdots$&$\cdots$&NRT&1484&55817--58852&194&4&512.0&61&50&$\textcolor{red}{4.2}^{9.0}_{\textcolor{red}{3.0}}$&$16^{40}_{9}$&$\textcolor{red}{44}^{\textcolor{red}{61}}_{21}$\\
$\cdots$&$\cdots$&JBO&1556&56167--58857&89&1&352.0&65&35&$2.3^{4.2}_{\textcolor{red}{1.3}}$&$13^{\textcolor{red}{27}}_{8}$&$80^{\textcolor{red}{142}}_{44}$\\
\\
J0636+5128&0.2&EFF&1347&56669--59021&23&1.56&200.0&10&28&$6.4^{37.5}_{3.4}$&$8.8^{\textcolor{red}{48.8}}_{4.6}$&$29^{55}_{5}$\\
$\cdots$&$\cdots$&JBO&1532&56655--58120&35&2&384.0&60&29&$9.7^{17.1}_{4.8}$&$8.0^{\textcolor{red}{32.6}}_{5.1}$&$19^{39}_{11}$\\
$\cdots$&$\cdots$&NRT&1544&56657--58419&97&4&392.0&61&60&$12^{40}_{\textcolor{red}{5}}$&$11^{\textcolor{red}{56}}_{5}$&$16^{\textcolor{red}{38}}_{5}$\\
\\
J1022+1001&0.4&NRT&1484&55839--58853&167&4&512.0&61&49&$64^{95}_{21}$&$\textcolor{red}{54}^{\textcolor{red}{125}}_{26}$&$2.9^{9.0}_{1.9}$\\
\\
J1600-3053&0.2&JBO&1532&56114--58810&45&1&400.0&60&38&$\textcolor{red}{1.1}^{2.0}_{\textcolor{red}{0.8}}$&$8.6^{13.8}_{6.5}$&$\textcolor{red}{160}^{\textcolor{red}{220}}_{90}$\\
$\cdots$&$\cdots$&NRT&2174&55872--56730&23&4&472.0&61&58&$\textcolor{red}{3.4}^{7.4}_{\textcolor{red}{2.6}}$&$15^{21}_{11}$&$\textcolor{red}{55}^{\textcolor{red}{71}}_{25}$\\
$\cdots$&$\cdots$&$\cdots$&2539&56807--58695&96&4&512.0&61&60&$6.5^{12.9}_{\textcolor{red}{3.6}}$&$16^{31}_{10}$&$28^{\textcolor{red}{52}}_{14}$\\
\\
J1640+2224&0.3&NRT&1484&55856--58819&107&4&512.0&61&58&$55^{92}_{20}$&$\textcolor{red}{89}^{\textcolor{red}{220}}_{39}$&$3.3^{9.1}_{2.0}$\\
\\
J1713+0747&0.4&EFF&1347&55652--58714&104&1.56&200.0&10&31&$15^{29}_{7}$&$\textcolor{red}{34}^{\textcolor{red}{80}}_{15}$&$13^{27}_{6}$\\
$\cdots$&$\cdots$&NRT&1510&55801--58269&315&4&460.0&61&44&$24^{57}_{12}$&$39^{\textcolor{red}{100}}_{18}$&$7.6^{15.8}_{3.2}$\\
$\cdots$&$\cdots$&$\cdots$&2565&56181--58113&73&4&460.0&61&55&$63^{99}_{30}$&$41^{\textcolor{red}{85}}_{21}$&$2.9^{6.2}_{1.9}$\\
\\
J1857+0943&0.2&EFF&1363&55634--58825&47&1.56&168.8&10&28&$4.7^{10.7}_{2.6}$&$25^{\textcolor{red}{37}}_{15}$&$39^{71}_{17}$\\
$\cdots$&$\cdots$&NRT&1510&55801--58838&131&4&460.0&61&56&$9.7^{20.7}_{\textcolor{red}{5.2}}$&$29^{\textcolor{red}{42}}_{20}$&$19^{\textcolor{red}{36}}_{9}$\\
$\cdots$&$\cdots$&JBO&1532&56256--58559&47&1&400.0&70&20&$9.4^{26.5}_{4.6}$&$21^{\textcolor{red}{39}}_{11}$&$20^{41}_{7}$\\
$\cdots$&$\cdots$&$\cdots$&2154&55866--57092&6&4&512.0&61&53&$22^{52}_{11}$&$31^{\textcolor{red}{51}}_{27}$&$8.7^{17.7}_{4.0}$\\
$\cdots$&$\cdots$&$\cdots$&2563&56860--57212&3&4&464.0&61&61&$63^{92}_{44}$&$58^{\textcolor{red}{73}}_{33}$&$2.9^{4.3}_{2.0}$\\
\\
J1939+2134&0.2&EFF&1349&55634--57734&74&1.56&140.6&10&29&$\textcolor{red}{1.2^{1.9}_{1.0}}$&$4.9^{6.7}_{3.3}$&$\textcolor{red}{150^{190}_{100}}$\\
$\cdots$&$\cdots$&WSRT&1395&54337--57196&144&0.3125&129.1&60&29&$0.8^{1.5}_{0.5}$&$5.9^{8.7}_{4.1}$&$230^{390}_{120}$\\
$\cdots$&$\cdots$&JBO&1566&55968--58557&146&1&332.0&116&30&$\textcolor{red}{1.4}^{2.4}_{\textcolor{red}{1.0}}$&$6.9^{8.8}_{5.3}$&$\textcolor{red}{130}^{\textcolor{red}{180}}_{80}$\\
$\cdots$&$\cdots$&NRT&2054&55805--57355&26&4&512.0&61&36&$\textcolor{red}{5.6}^{10.7}_{\textcolor{red}{3.7}}$&$9.6^{13.2}_{6.8}$&$\textcolor{red}{33}^{\textcolor{red}{50}}_{17}$\\
$\cdots$&$\cdots$&$\cdots$&2154&56510--57574&6&4&512.0&61&20&$\textcolor{red}{5.9}^{7.3}_{\textcolor{red}{3.8}}$&$11^{13}_{8}$&$\textcolor{red}{32}^{\textcolor{red}{49}}_{25}$\\
$\cdots$&$\cdots$&$\cdots$&2533&56707--58235&29&4&412.0&61&31&$6.6^{12.3}_{\textcolor{red}{4.8}}$&$10^{14}_{7}$&$28^{\textcolor{red}{39}}_{15}$\\
$\cdots$&$\cdots$&EFF&2634&55632--57229&42&1.56&112.5&10&29&$5.7^{12.4}_{\textcolor{red}{2.2}}$&$7.6^{16.6}_{4.1}$&$33^{\textcolor{red}{85}}_{15}$\\

\\
J2145-0750&0.4&NRT&1550&55804--58269&195&4&380.0&61&52&$34^{74}_{17}$&$32^{\textcolor{red}{80}}_{16}$&$5.4^{10.8}_{2.5}$\\
\\
J2214+3000&0.2&NRT&1484&55828--58085&41&4&512.0&61&61&$24^{57}_{13}$&$21^{41}_{12}$&$7.8^{14.1}_{3.2}$\\
\\
J2234+0944&0.3&NRT&1484&55825--58150&88&4&512.0&61&51&$41^{76}_{22}$&$32^{\textcolor{red}{58}}_{18}$&$4.5^{8.6}_{2.4}$\\
\\
J2317+1439&0.3&NRT&1484&55877--58841&182&4&512.0&61&47&$39^{74}_{20}$&$34^{\textcolor{red}{72}}_{17}$&$4.8^{9.1}_{2.5}$\\
\\
	\hline
	\hline
  \end{tabular}
\tablefoot{Where $f_d$ is the filling factor in Equation \ref{eq:error_sys}, $f_c$ is the effective center frequency, N$_{\rm{obs}}$ is the number of observations, CHBW is the channel bandwidth, BW is the effective bandwidth of the observation after the removal of RFI, t$_{\rm{sub}}$ is the subintegration length, $\bar{\rm{t}}_{\rm{obs}}$ is the mean observation length. $\nu_{\rm d}$, $\tau_{\rm d}$ and $\tau_{\rm{st}}$ are the scintillation bandwidth, the scintillation timescale and the scattering timescale, respectively, where we present their median and their 5/95 percentiles as sub/superscripts. Values in red indicate that measurements are not reliable and are to be considered as upper (for $\nu_{\rm d}$) or lower (for $\tau_{\rm d}$) limits. All values and uncertainties have been rounded to the least significant digit shown.}
\end{table*}

\begin{figure*}
\centering
\includegraphics[width=16cm]{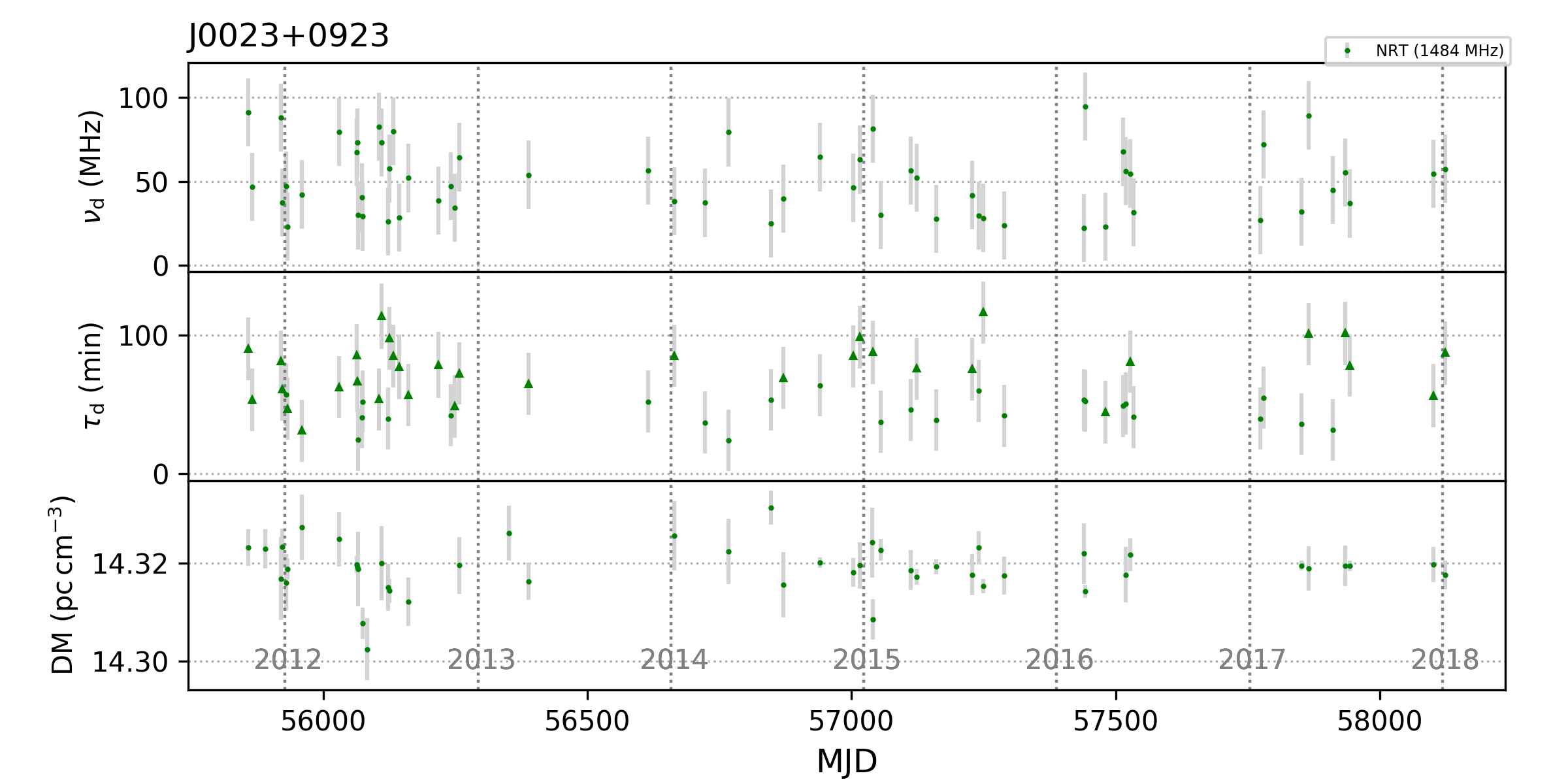}
\includegraphics[width=16cm]{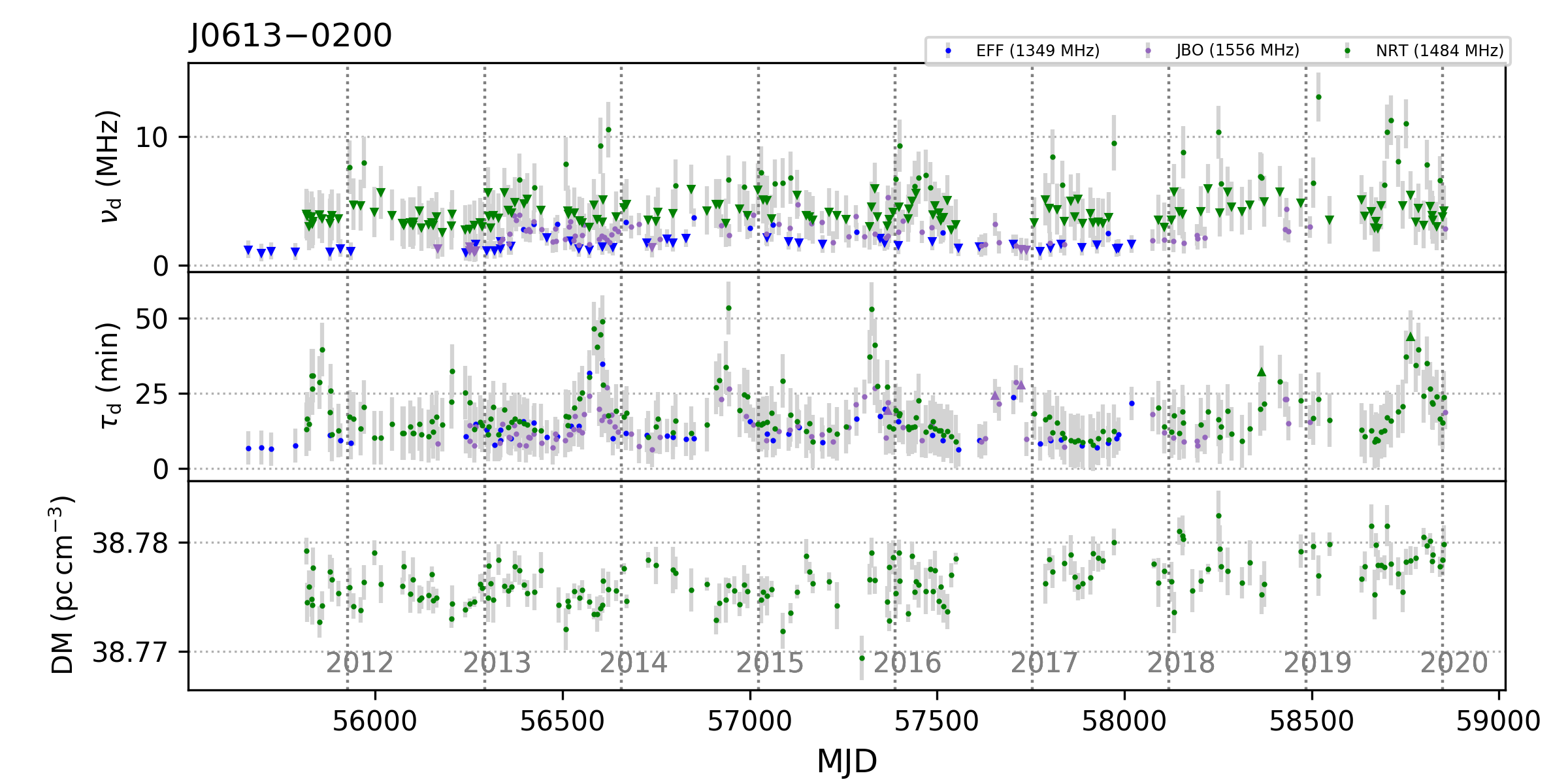}
\includegraphics[width=16cm]{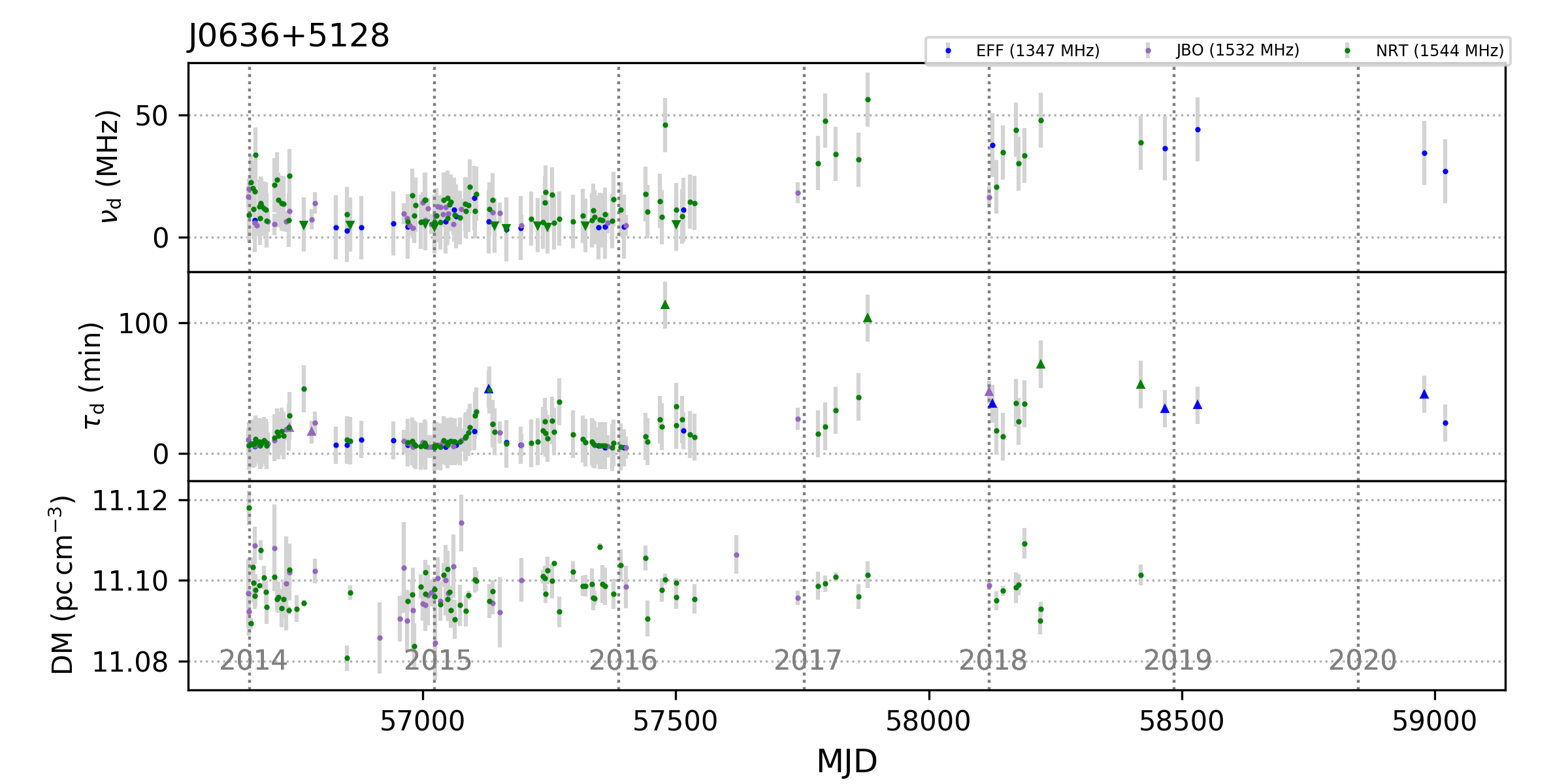}
\end{figure*}

\begin{figure*}
\centering
\includegraphics[width=16cm]{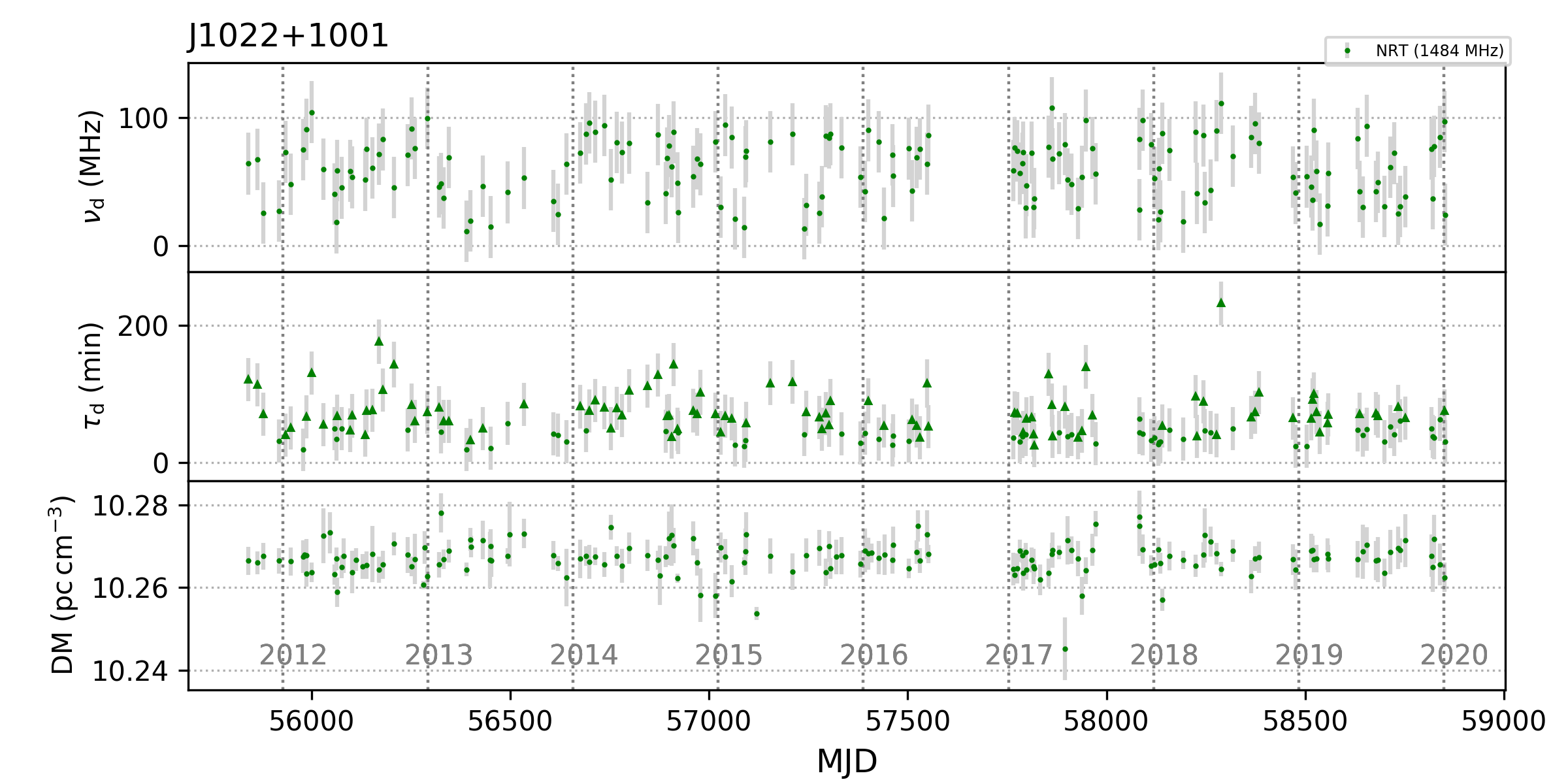}
\includegraphics[width=16cm]{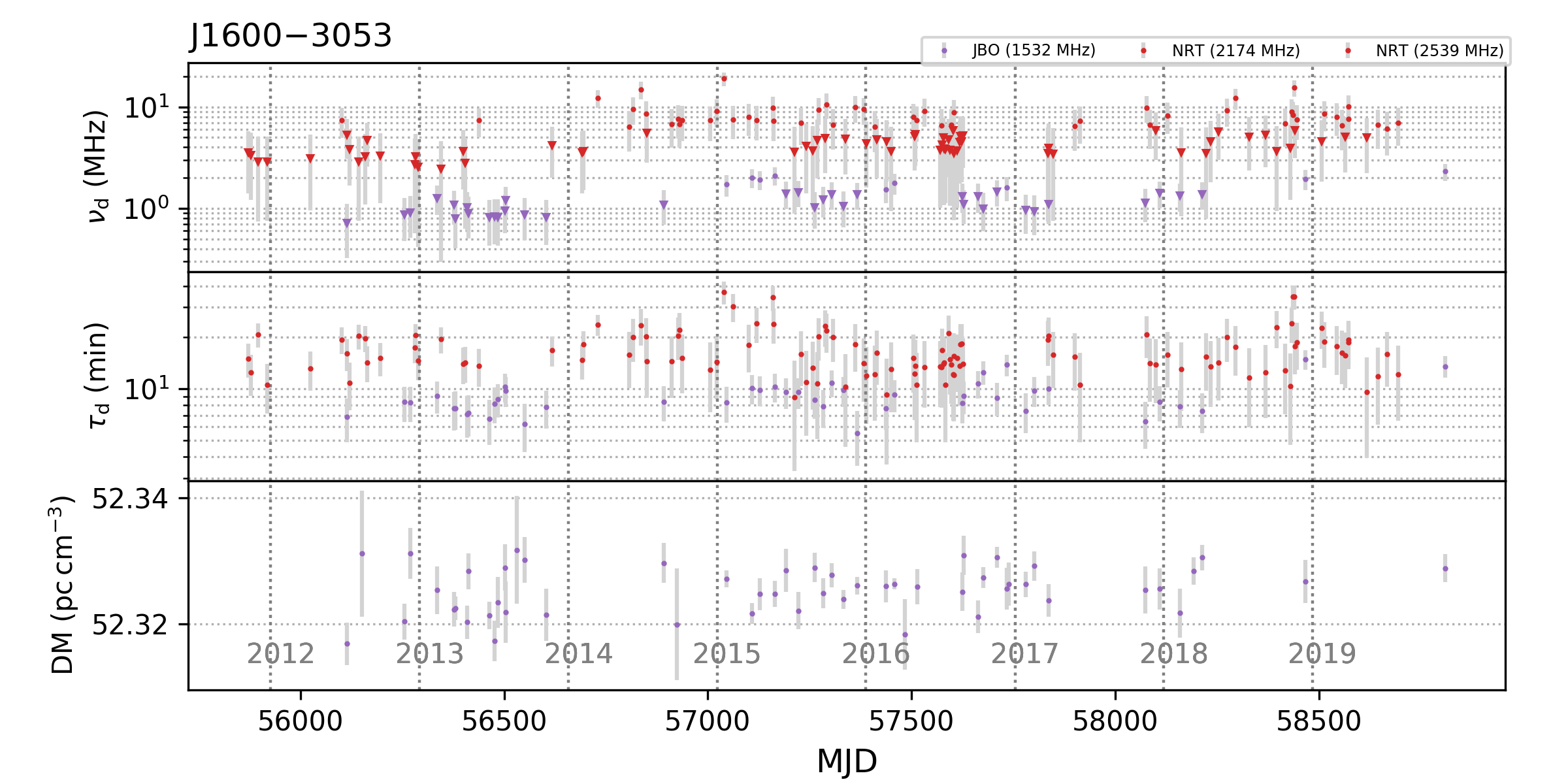}
\includegraphics[width=16cm]{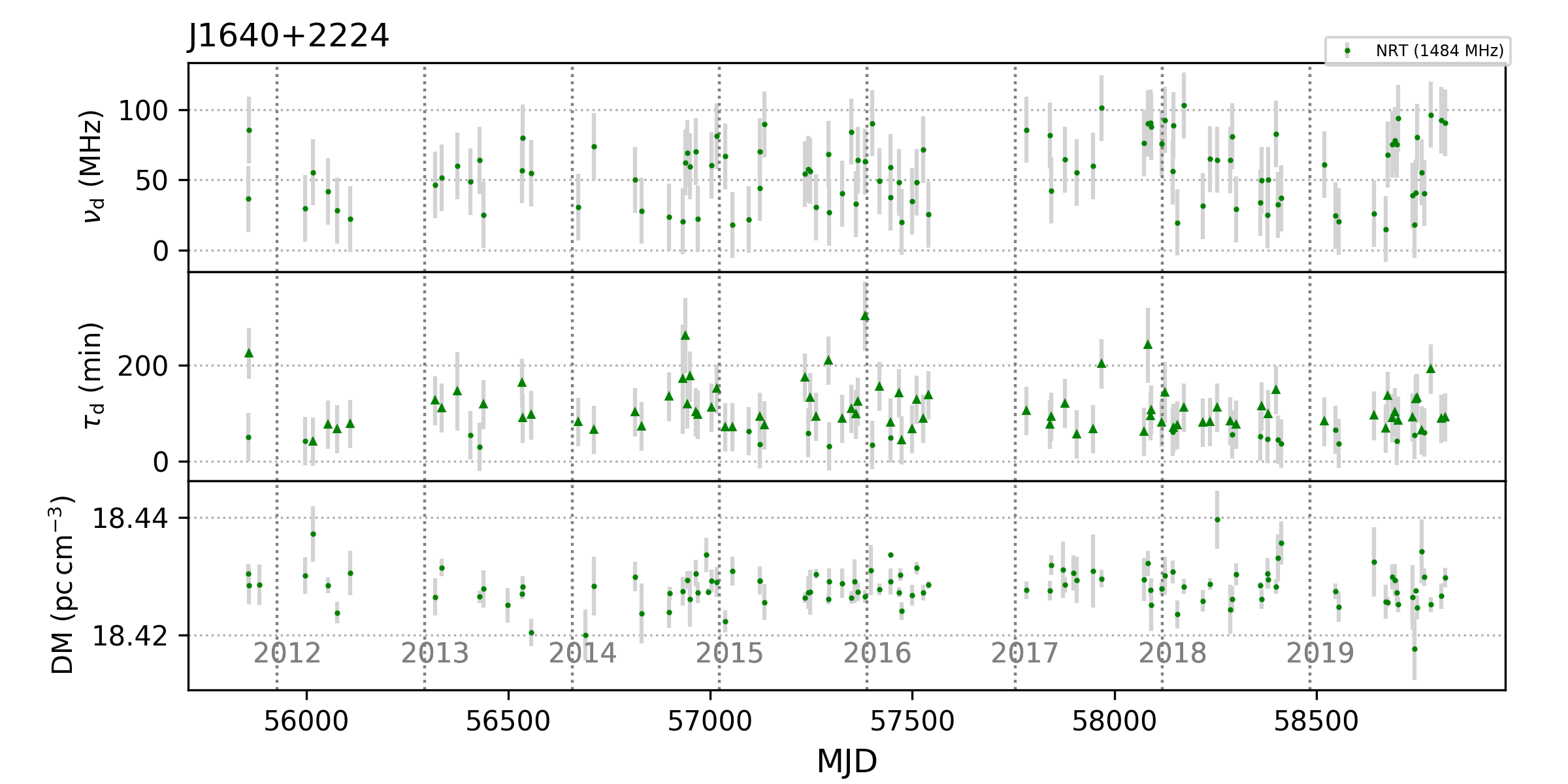}
\end{figure*}

\begin{figure*}
\centering
\includegraphics[width=16cm]{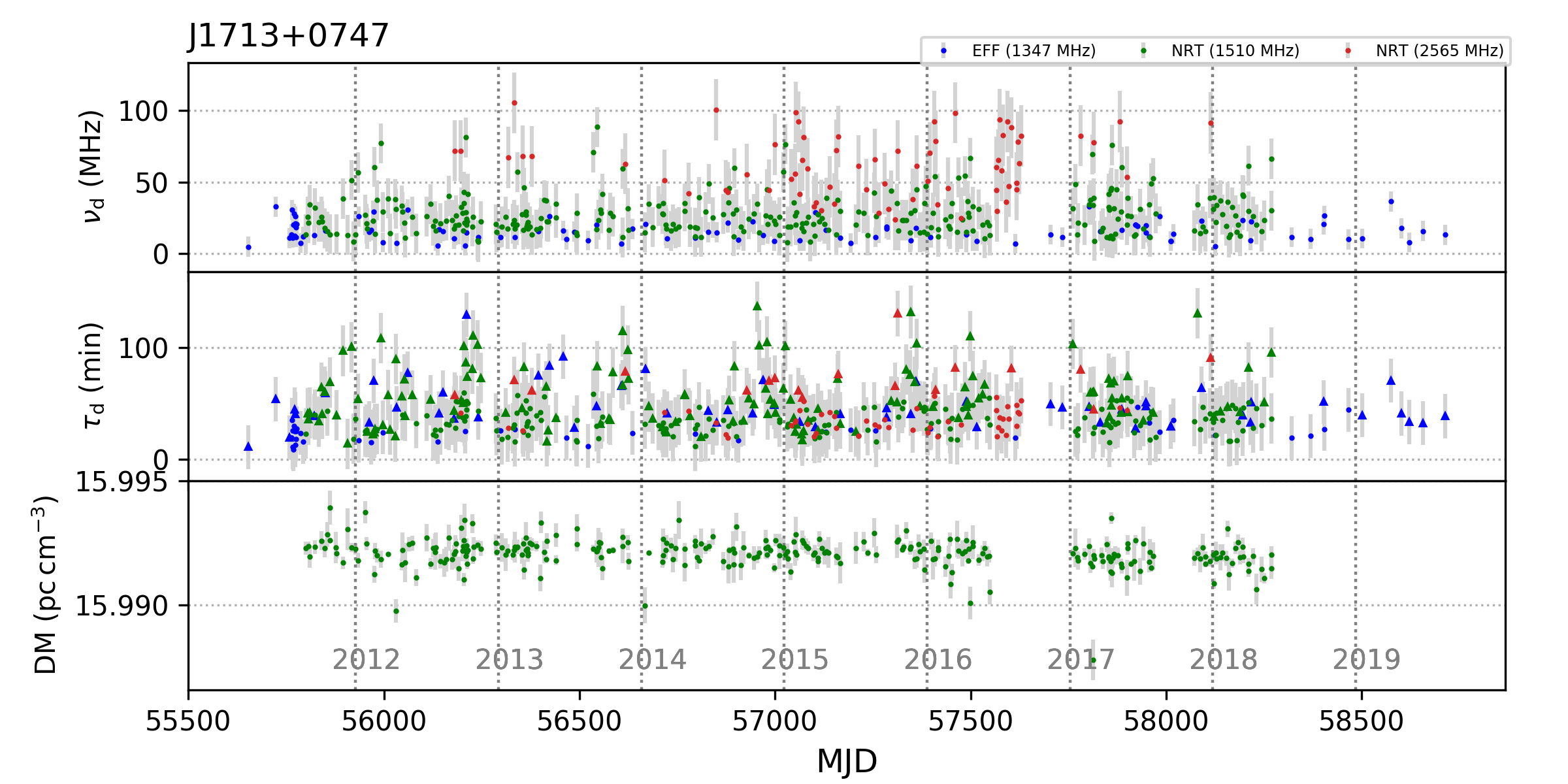}
\includegraphics[width=16cm]{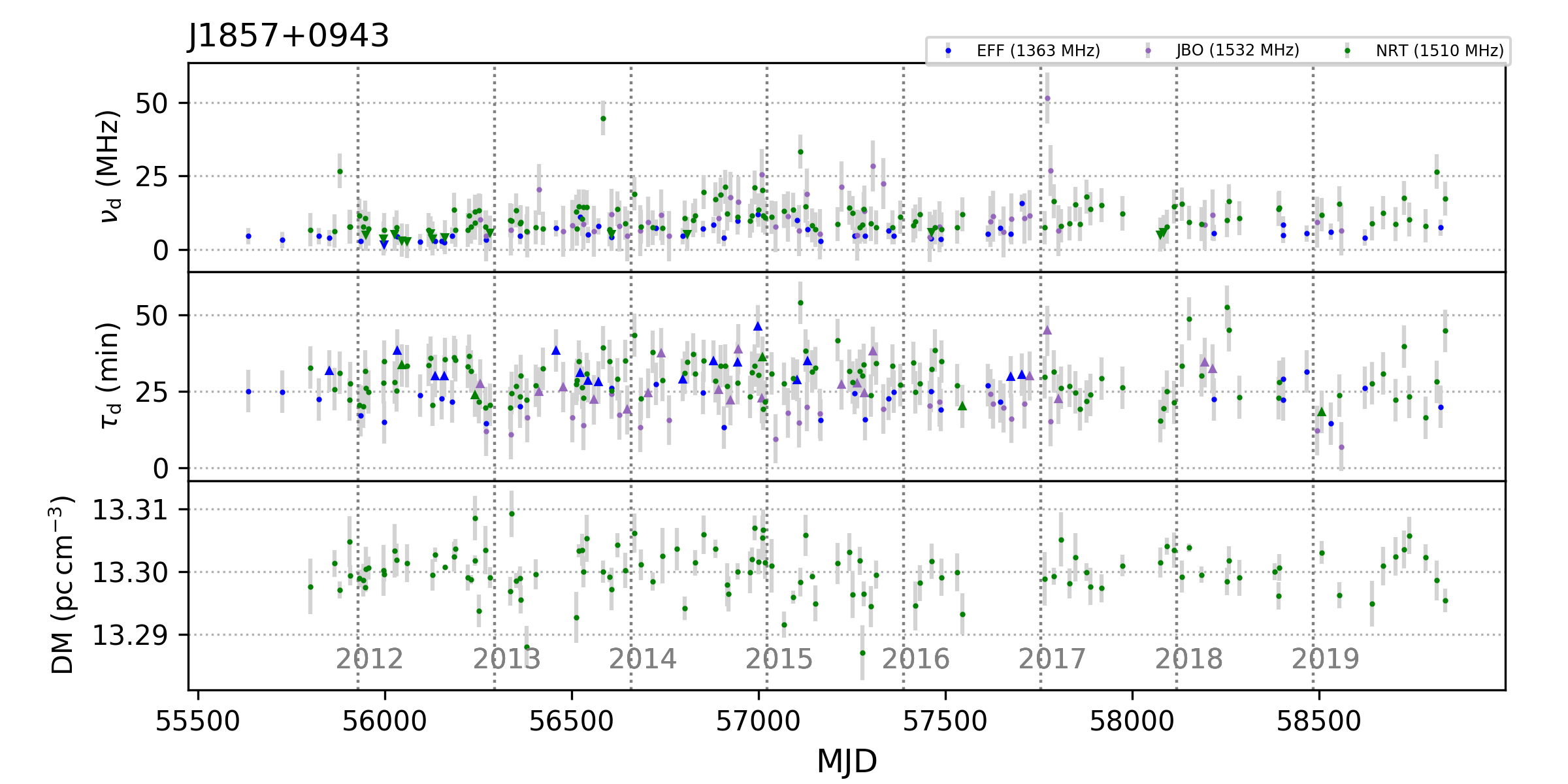}
\includegraphics[width=16cm]{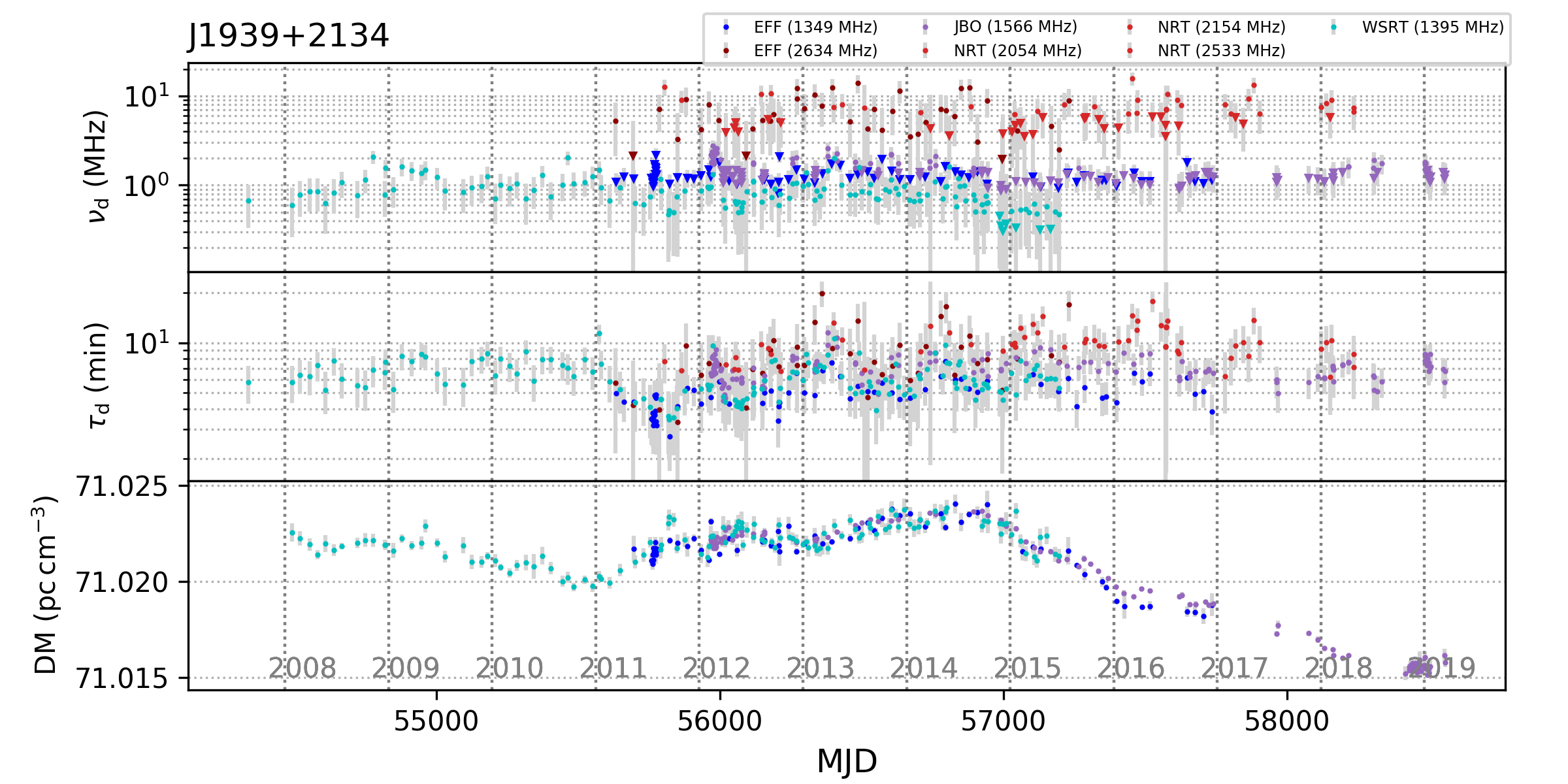}
\end{figure*}

\begin{figure*}
\centering
\includegraphics[width=16cm]{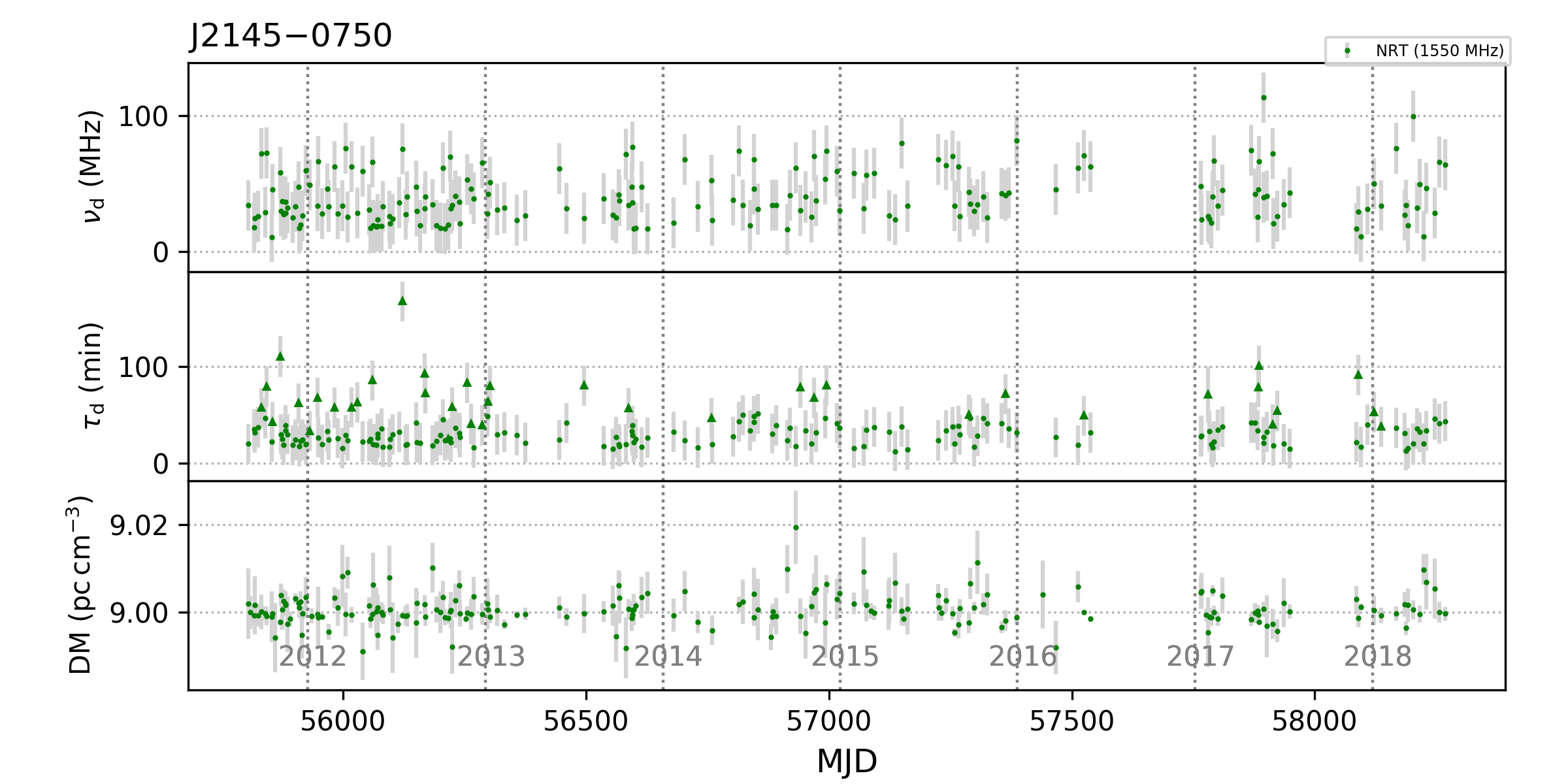}
\includegraphics[width=16cm]{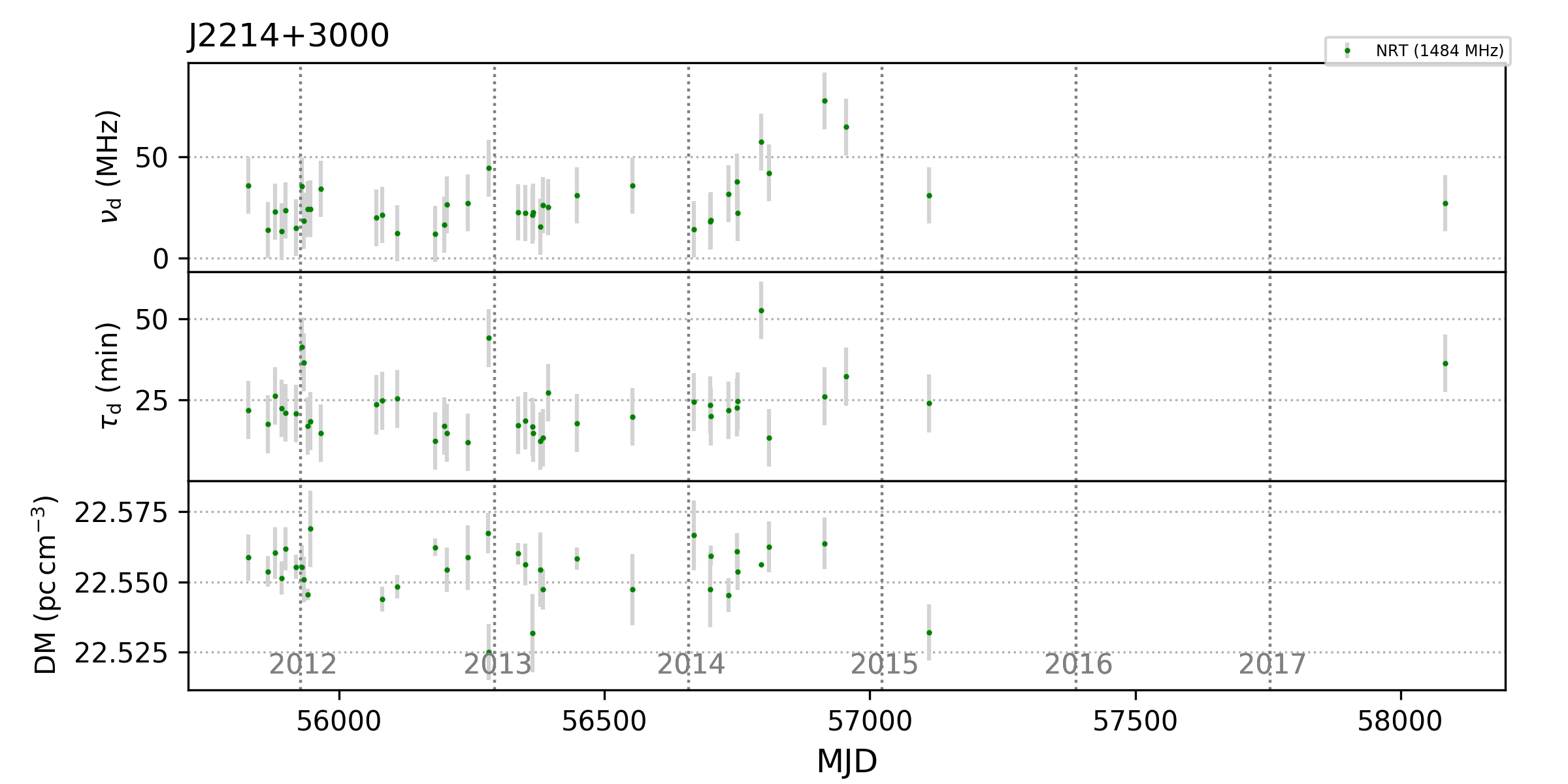}
\includegraphics[width=16cm]{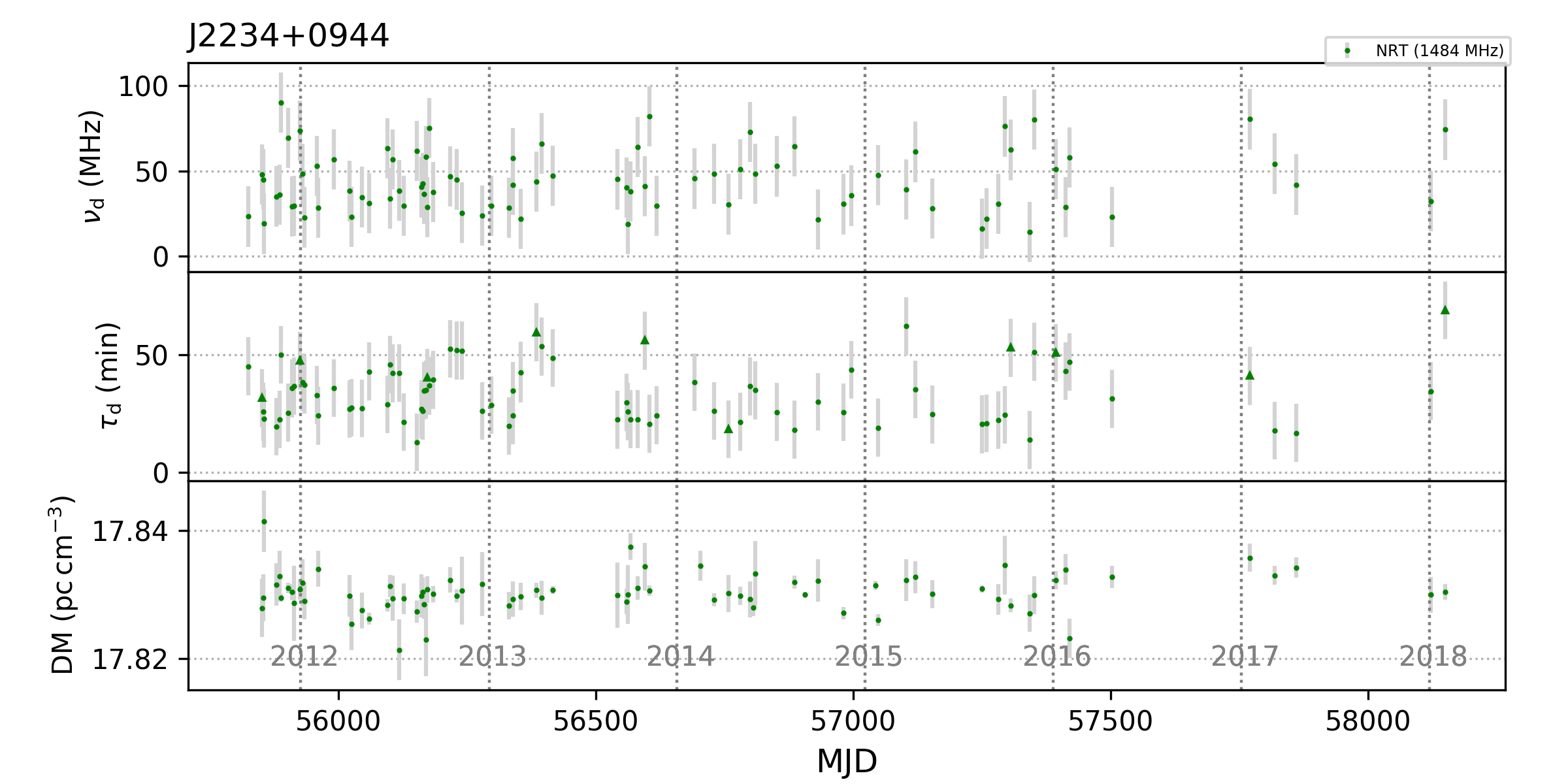}
\end{figure*}

\begin{figure*}
\centering
\includegraphics[width=16cm]{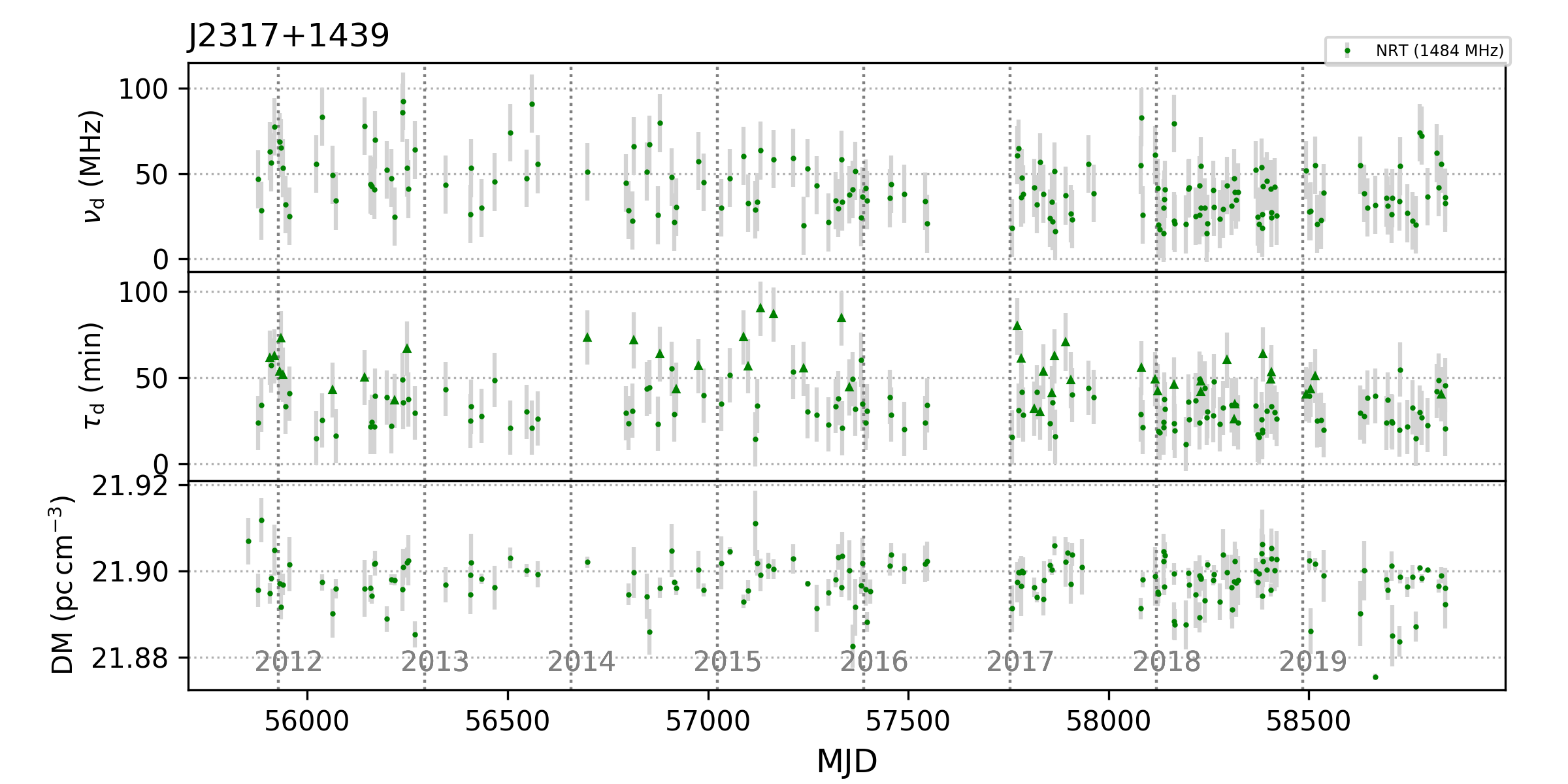}
\caption{Long-term variations of scintillation bandwidth $\nu_{\rm d}$, scintillation timescale $\tau_{\rm d}$ and DM for 13 EPTA pulsars. To ensure that the variations shown by the most precise DM values remain visible, for most pulsars, we only show the most precise DM measurements, which are usually derived from 21-cm Nan{\c c}ay data. The different colors indicate  measurements at different telescopes, following the legend given in the top-right corner of each sub plot. Following the telescope name, we present the center frequency in units of MHz. 
Downward-pointing and upward-pointing triangles indicate unreliable estimates (i.e., upper and lower limits) caused by limited frequency resolution and limited observation length, respectively. The vertical gray dotted lines indicate the start of a calendar year.}
\label{fig:TS}
\end{figure*}
      
According to the method described in Section~\ref{sec:methods}, we present the median of $\nu_{\rm d}$ and $\tau_{\rm d}$ for 13 pulsars in columns 11 and 12 of Table~\ref{tab:scinpara}, the sub and super scripts are the 5/95 percentiles. To analyze the long-term variations of pulsar scintillation, we plotted the time series of $\nu_{\rm d}$ and $\tau_{\rm d}$  of all pulsars in the first and second panels of Figure~\ref{fig:TS}, respectively. The time series of DM for each pulsar was plotted in lowest panels of Figure~\ref{fig:TS}.
Some pulsars have been observed at both 21-cm and 11-cm bands, in which case the significant difference in scintillation parameters causes the 21-cm values to be compressed at the bottom of the scale, with any variations invisible. To remedy this, for pulsars with measurements at different bands, the scintillation parameters have been plotted on a logarithmic scale. 
If the number of observations in a given observing band is too small for an analysis for the variation of scintillation parameters, we did not plot the measurements from this observing band in Figure~\ref{fig:TS}, for example, the measurements of PSR~J1857+0943 at 11-cm band. The data for all our measurements are publicly available on Zenodo at \url{https://sandbox.zenodo.org/record/1016820#.Yg5VhC8w2Td}\footnote{The uncertainties of the scintillation parameters we presented in Zenodo consist of the formal uncertainties from the fitting procedure and the statistical uncertainties due to the finite number of scintles in the dynamic spectrum, the additional uncertainties to reflect the Gaussian variations of measurements have not been included.}. 

In addition to the measured scintillation parameters, the relevant observation information is also listed in Table~\ref{tab:scinpara}, including the effective center frequency $f_c$, the number of observations N$_{\rm{obs}}$, the channel bandwidth (CHBW), the effective bandwidth (BW), the subintegration length t$_{\rm{sub}}$ and the mean observation length $\bar{\rm{t}}_{\rm{obs}}$. The effective center frequencies and the effective bandwidths in Table~\ref{tab:scinpara} vary slightly as described in Section~\ref{sec:data} because edge frequency channels were sometimes removed due to RFI.

Observations for some pulsars at some telescopes could not be used to derive scintillation parameters and were therefore not included in the results presented in Table~\ref{tab:scinpara} and Figure~\ref{fig:TS}.  We describe the specific details of which observations were not suitable below: (i) due to limited frequency resolution, some data sets were not able to resolve any scintles in the dynamic spectrum, for examples, WSRT observations of PSR~J0636+5128 at a center frequency of 346.25\,MHz and NRT observations of PSR~J1600$-$3053 at a center frequency of 1484\,MHz; (ii) data sets with narrow observing bandwidth and weaker scintillation, where one full scintle is larger than the available frequency range, for examples, EFF observations of PSR~J2145$-$0750 at a center frequency of 2627\,MHz and NRT observavtions of PSR~1713+0747 at a center frequency of 2539\,MHz; and (iii) some observations showed a lot of RFI and the dynamic spectrum was badly damaged after RFI removal, which seriously affects the derived scintillation parameters.
If more than half of a set of observations are accompanied by severe RFI, we abandoned all observations from that set, for example, JBO observations of PSR~J2317+1439 at a center frequency of 1532\,MHz. If only a few observations in a set showed extensive RFI, we just removed those observations.

As shown in the first panel of Figure~\ref{fig:dyacf}, in some observations, the scintles are not fully resolved by the frequency resolution in the dynamic spectrum. For those observations, usually, the 1D ACF has less than three points within the half-power width, which leads to an unreliable (usually overestimated) measurement of $\nu_{\rm d}$. Besides, we found that almost all $\nu_{\rm d}$ of those observations are smaller than 1.5 times the channel bandwidth.
Thus, if the value we derived for $\nu_{\rm d}$ is smaller than 1.5 times the channel bandwidth, we considered this value to be unreliable, use it as an upper limit and use a downward-pointing triangle to plot it in Figure~\ref{fig:TS}.
If, in a given data set, 5\% or more of the $\nu_{\rm d}$ measurements are upper limits, then the 5th percentile is unreliable, hence the subscript is shown in red in Table~\ref{tab:scinpara}; similarly, the median value for $\nu_{\rm d}$ is shown in red if 50\% or more of measurements are upper limits; and the 95th percentile (superscript) is shown in red if more than 95\% of the measurements are upper limits.

Some observations can not show one full scintle in the time axis of the dynamic spectrum because of the limited observation duration, as shown in the fourth plot of Figure~\ref{fig:dyacf}: in this case the minimum value of the ACF along the time axis is larger than 1/e.
For those observations, we fit all points of the time axis and extrapolated the result to the delay where the time lag is smaller than or equal to 1/e, and then we used the half-width at 1/e along this virtually extended time axis of the ACF as the value for $\tau_{\rm d}$. However, the $\tau_{\rm d}$ estimated in this way is larger than the observing duration $\rm{t_{obs}}$ and is unreliable, so we used an upward-pointing triangle to plot them in Figure~\ref{fig:TS}. 
Similarly to the treatment of $\nu_{\rm d}$, values for $\tau_{\rm d}$ are shown in red in Table 4, depending on what fraction of measurements were considered to be limited by the observing length of the data. Contrary to the treatment of $\nu_{\rm d}$, we sorted $\tau_{\rm d}$ in descending order and we marked the 5th percentile as the superscript (the 95th percentile as the subscript).

\section{Comparison}
\label{sec:comparison}
In this section, we compare the results presented in Table~\ref{tab:scinpara} first against previously published observational results and subsequently against the theoretical predictions of the NE2001 Galactic electron density model.

\begin{table*}\tiny
\renewcommand\arraystretch{1.4}
\centering
\caption{Scintillation parameters from previous publications and NE2001 electron-density model predictions}
  \label{tab:scincompare} 
  \begin{tabular}{ccc|cccc|ccc}
	\hline 
	\hline
 & &   & \multicolumn{4}{c|}{$\nu_{d}$ (MHz) }  & \multicolumn{3}{c}{$\tau_{d}$ (min)}\\
Pulsar & DM & $f_{\rm c}$, BW & This work& \citet{lmj16}& \citet{tmc+21} & NE2001 & This work & \citet{tmc+21} & NE2001\\
 & pc\,cm$^{-3}$ & (MHz) & 2011-2020  & 2010-2013 & 2011-2017&  &2011-2020 &2011-2017&  \\
\hline
J0023+0923&14.3&1484(512)&$47^{88}_{24}$&$20^{26}_{13}$&$\ast$&$71^{143}_{31}$&$\textcolor{red}{57}^{\textcolor{red}{101}}_{32}$&$\ast$&$28^{33}_{22}$\\
&&&&&&&&&\\ 
 J0613-0200&38.8&1349(141)&$\textcolor{red}{1.7}^{3.3}_{\textcolor{red}{1.1}}$&$7.0^{9.7}_{4.3}$&$2.5^{4.4}_{0.6}$&$7.1^{8.9}_{5.6}$&$11^{24}_{7}$&$8.8^{12.3}_{5.3}$&$20^{21}_{19}$\\
 $\cdots$&$\cdots$&1484(512)&$\textcolor{red}{4.2}^{9.0}_{\textcolor{red}{3.0}}$&$11^{15}_{6}$&$3.8^{6.7}_{1.0}$&$11^{22}_{5}$&$16^{40}_{9}$&$9.9^{13.8}_{5.9}$&$22^{27}_{18}$\\
$\cdots$&$\cdots$&1556(352)&$2.3^{4.2}_{\textcolor{red}{1.3}}$&$13^{18}_{8}$&$4.7^{8.2}_{1.2}$&$13^{21}_{8}$&$13^{\textcolor{red}{27}}_{8}$&$10^{15}_{6}$&$24^{27}_{21}$\\
&&&&&&&&&\\ 
 J0636+5128&11.1&1347(200)&$6.4^{37.5}_{3.4}$&$\ast$&$4.4^{6.9}_{1.9}$&$61^{83}_{43}$&$8.8^{\textcolor{red}{48.8}}_{4.6}$&$7.9^{10.5}_{5.3}$&$380^{410}_{340}$\\
$\cdots$&$\cdots$&1532(384)&$9.7^{17.1}_{4.8}$&$\ast$&$7.7^{12.1}_{3.3}$&$110^{180}_{60}$&$8.0^{\textcolor{red}{32.6}}_{5.1}$&$9.2^{12.3}_{6.2}$&$440^{510}_{370}$\\
$\cdots$&$\cdots$&1544(392)&$12^{40}_{\textcolor{red}{5}}$&$\ast$&$7.9^{12.5}_{3.4}$&$110^{190}_{60}$&$11^{\textcolor{red}{56}}_{5}$&$9.3^{12.4}_{6.2}$&$440^{510}_{380}$\\
&&&&&&&&&\\ 
 J1022+1001&10.3&1484(512)&$64^{95}_{21}$&$\ast$&$\ast$&$130^{260}_{60}$&$\textcolor{red}{54}^{\textcolor{red}{125}}_{26}$&$\ast$&$40^{48}_{32}$\\
&&&&&&&&&\\ 
 J1600-3053&52.3&1532(400)&$\textcolor{red}{1.1}^{2.0}_{\textcolor{red}{0.8}}$&$\ast$&$\ast$&$2.2^{3.7}_{1.2}$&$8.6^{13.8}_{6.5}$&$\ast$&$5.2^{6.0}_{4.4}$\\
$\cdots$&$\cdots$&2174(472)&$\textcolor{red}{3.4}^{7.4}_{\textcolor{red}{2.6}}$&$\ast$&$\ast$&$10^{16}_{6}$&$15^{21}_{11}$&$\ast$&$7.9^{8.9}_{6.9}$\\
$\cdots$&$\cdots$&2539(512)&$6.5^{12.9}_{\textcolor{red}{3.6}}$&$\ast$&$\ast$&$20^{31}_{13}$&$16^{31}_{10}$&$\ast$&$9.5^{10.7}_{8.4}$\\
&&&&&&&&&\\ 
 J1640+2224&18.4&1484(512)&$55^{92}_{20}$&$54^{68}_{40}$&$48^{72}_{23}$&$30^{61}_{13}$&$\textcolor{red}{89}^{\textcolor{red}{220}}_{39}$&$>30$&$21^{25}_{16}$\\
&&&&&&&&&\\ 
 J1713+0747&16.0&1347(200)&$15^{29}_{7}$&$13^{19}_{8}$&$14^{24}_{5}$&$28^{39}_{20}$&$\textcolor{red}{34}^{\textcolor{red}{80}}_{15}$&$>30$&$51^{55}_{46}$\\
$\cdots$&$\cdots$&1510(460)&$24^{57}_{12}$&$22^{31}_{13}$&$24^{39}_{8}$&$47^{87}_{22}$&$39^{\textcolor{red}{100}}_{18}$&$>30$&$58^{69}_{48}$\\
$\cdots$&$\cdots$&2565(460)&$63^{99}_{30}$&$220^{310}_{130}$&$240^{400}_{80}$&$480^{700}_{320}$&$41^{\textcolor{red}{85}}_{21}$&$>30$&$110^{120}_{100}$\\
&&&&&&&&&\\ 
 J1857+0943&13.3&1363(169)&$4.7^{10.7}_{2.6}$&$3.4^{5.0}_{1.8}$&$6.6^{9.8}_{3.3}$&$30^{39}_{23}$&$25^{\textcolor{red}{37}}_{15}$&$>30$&$54^{58}_{50}$\\
$\cdots$&$\cdots$&1510(460)&$9.7^{20.7}_{\textcolor{red}{5.2}}$&$5.4^{7.8}_{2.9}$&$10^{15}_{5}$&$48^{89}_{23}$&$29^{\textcolor{red}{42}}_{20}$&$>30$&$61^{72}_{50}$\\
$\cdots$&$\cdots$&1532(400)&$9.4^{26.5}_{4.6}$&$5.7^{8.3}_{3.1}$&$11^{16}_{5}$&$51^{87}_{27}$&$21^{\textcolor{red}{39}}_{11}$&$>30$&$62^{72}_{52}$\\
$\cdots$&$\cdots$&2154(512)&$22^{52}_{11}$&$26^{37}_{14}$&$49^{74}_{25}$&$230^{370}_{130}$&$31^{\textcolor{red}{51}}_{27}$&$>30$&$93^{106}_{80}$\\
$\cdots$&$\cdots$&2563(464)&$63^{92}_{44}$&$55^{80}_{30}$&$110^{160}_{50}$&$490^{710}_{320}$&$58^{\textcolor{red}{73}}_{33}$&$>30$&$110^{130}_{100}$\\
&&&&&&&&&\\ 
 J1939+2134&71.0&1349(141)&$\textcolor{red}{1.2^{1.9}_{1.0}}$&$1.8^{2.6}_{0.9}$&$0.9^{1.4}_{0.4}$&$0.9^{1.1}_{0.7}$&$4.9^{6.7}_{3.3}$&$7.9^{10.6}_{5.3}$&$58^{62}_{54}$\\
 $\cdots$&$\cdots$&1395(129)&$0.8^{1.5}_{0.5}$&$2.0^{3.0}_{1.1}$&$1.1^{1.7}_{0.5}$&$1.0^{1.3}_{0.8}$&$5.9^{8.7}_{4.1}$&$8.2^{11.0}_{5.5}$&$60^{64}_{57}$\\
 $\cdots$&$\cdots$&1566(332)&$\textcolor{red}{1.4}^{2.4}_{\textcolor{red}{1.0}}$&$3.4^{5.0}_{1.8}$&$1.8^{2.8}_{0.8}$&$1.7^{2.7}_{1.0}$&$6.9^{8.8}_{5.3}$&$9.5^{12.6}_{6.3}$&$69^{78}_{61}$\\
$\cdots$&$\cdots$&2054(512)&$\textcolor{red}{5.6}^{10.7}_{\textcolor{red}{3.7}}$&$11^{16}_{6}$&$6.0^{9.2}_{2.8}$&$5.6^{9.5}_{3.1}$&$9.6^{13.2}_{6.8}$&$13^{17}_{9}$&$96^{111}_{82}$\\
$\cdots$&$\cdots$&2154(512)&$\textcolor{red}{5.9}^{7.3}_{\textcolor{red}{3.8}}$&$14^{20}_{7}$&$7.4^{11.3}_{3.4}$&$7.0^{11.4}_{4.0}$&$11^{13}_{8}$&$14^{19}_{9}$&$100^{120}_{90}$\\
$\cdots$&$\cdots$&2533(412)&$6.6^{12.3}_{\textcolor{red}{4.8}}$&$28^{41}_{15}$&$15^{23}_{7}$&$14^{20}_{10}$&$10^{14}_{7}$&$17^{23}_{11}$&$120^{140}_{110}$\\
$\cdots$&$\cdots$&2634(113)&$5.7^{12.4}_{\textcolor{red}{2.2}}$&$33^{49}_{18}$&$18^{27}_{8}$&$17^{19}_{15}$&$7.6^{16.6}_{4.1}$&$18^{24}_{12}$&$130^{130}_{130}$\\
&&&&&&&&&\\ 
 J2145-0750&9.0&1550(380)&$34^{74}_{17}$&$55^{71}_{40}$&$50^{62}_{37}$&$400^{670}_{230}$&$32^{\textcolor{red}{80}}_{16}$&$>30$&$110^{120}_{90}$\\
&&&&&&&&&\\ 
 J2214+3000&22.6&1484(512)&$24^{57}_{13}$&$22^{39}_{5}$&$\ast$&$56^{112}_{24}$&$21^{41}_{12}$&$\ast$&$68^{82}_{54}$\\
&&&&&&&&&\\ 
 J2234+0944&17.8&1484(512)&$41^{76}_{22}$&$\ast$&$\ast$&$52^{105}_{23}$&$32^{\textcolor{red}{58}}_{18}$&$\ast$&$17^{20}_{13}$\\
&&&&&&&&&\\ 
 J2317+1439&21.9&1484(512)&$39^{74}_{20}$&$40^{51}_{29}$&$44^{57}_{31}$&$91^{184}_{40}$&$34^{\textcolor{red}{72}}_{17}$&$>30$&$77^{94}_{62}$\\
\\
	\hline
	\hline
  \end{tabular}
\tablefoot{The values of this work come from Table~\ref{tab:scinpara}. $\nu_{\rm d}$ and $\tau_{\rm d}$ of the previous studies and the NE2001 simulations are rescaled to our observing frequencies using a scaling index of 4.4 and 1.2, respectively. Besides, as the NE2001 Electron Density Model assumes that all pulsars have a pulsar transverse velocity of 100\,km\,s$^{-1}$, we also used the estimated pulsar transverse velocities, that are calculated by using Equation~\ref{eq:vp} and are listed in Table~\ref{tab:velocity}, to adjust the scintillation timescale $\tau_d$ for each pulsar, like Equation (46) of \citet{cl91}.
The subscripts and superscripts for the previously published scintillation parameters were obtained by adding and subtracting the uncertainties first from the literature and then rescaling. 
For the NE2001 predictions, the subscripts and superscripts are the predicted values at the lowest and highest frequencies of our bandwidth, respectively.  See Section~\ref{sec:comparison} for a discussion of the reasons for discrepancies between this work, previous publications and NE2001.}
\end{table*}

\subsection{Comparison with previous studies}
\begin{figure*}
\centering
\includegraphics[width=9cm]{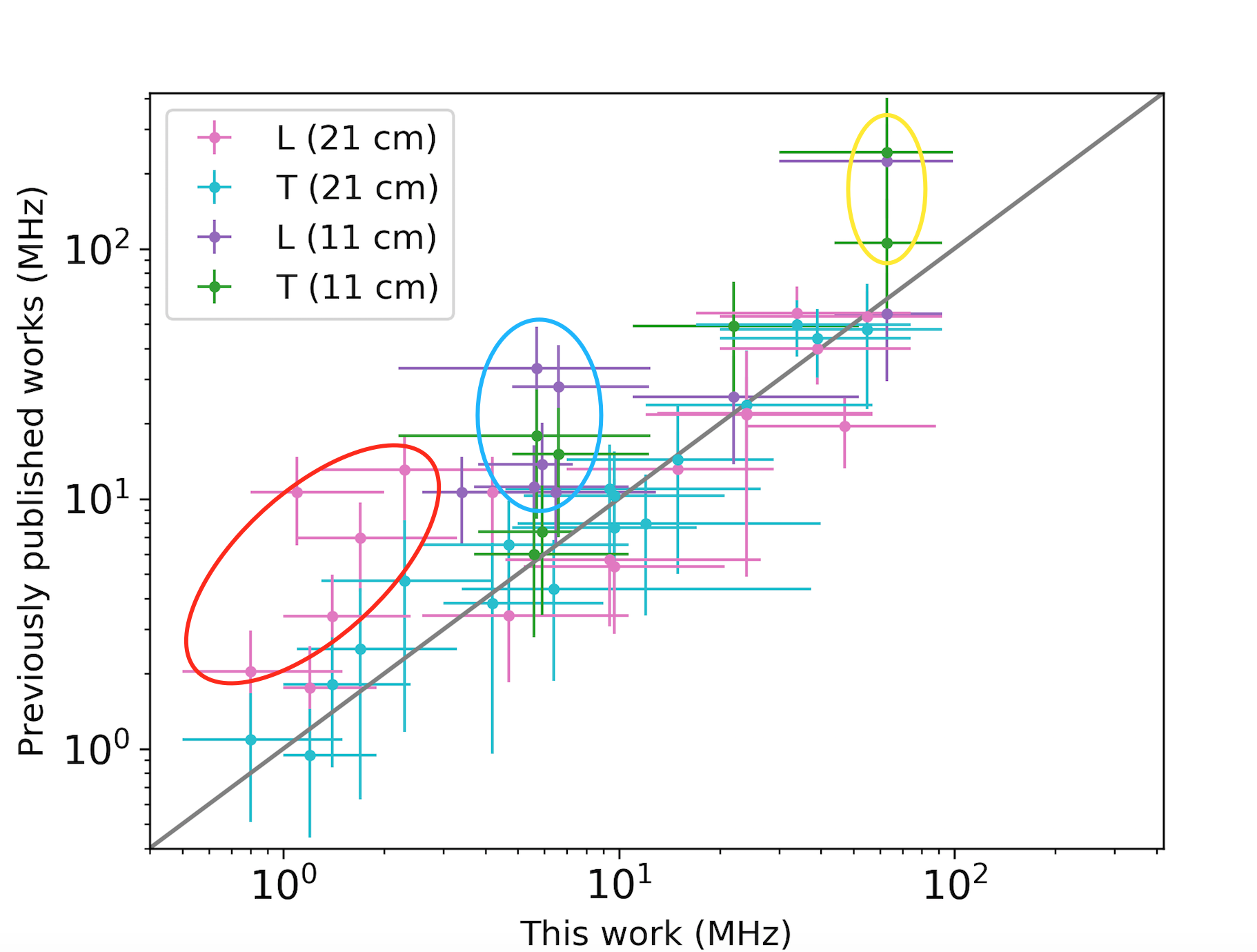}
\includegraphics[width=9.2cm]{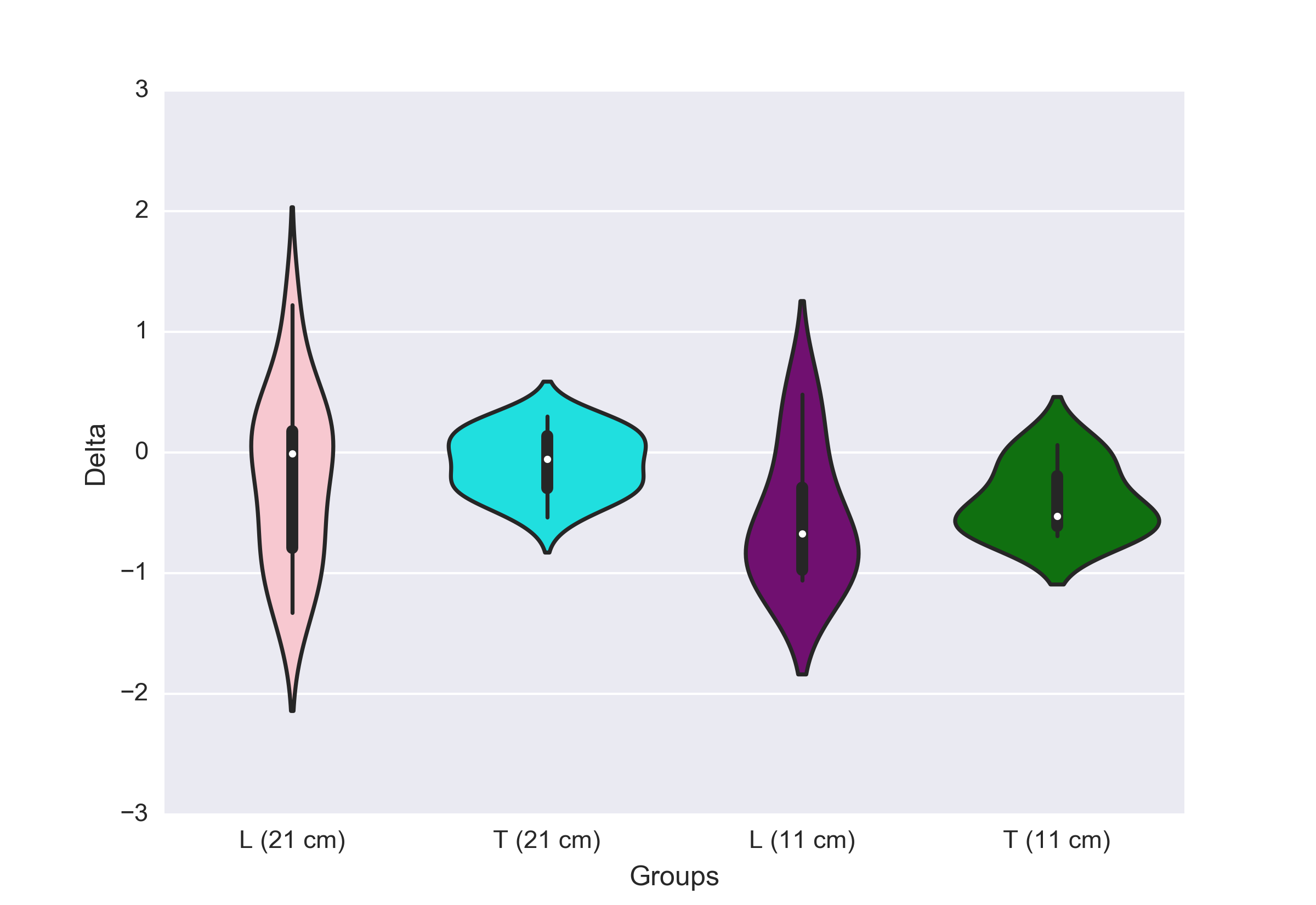}
\caption{The left panel compares the $\nu_{\rm d}$ from previous works to our measurements. Here "L" indicates \citet{lmj16}, "T" refers to \citet{tmc+21}, "11cm" compares against measurements we made in the 2-GHz range; "21cm" refers to measurements we made in the 1-GHz range -- in both cases the previously published measurements were rescaled to our observing frequency, as given in column 3 of Table~\ref{tab:scincompare}. Data from \citet{lmj16} were at 1500 MHz while \citet{tmc+21} used data at 1500 MHz as well as 820 MHz, but we only compare to the former. The three ellipses indicate groups of points that show significant offsets, which are discussed in more detail in the text.  
The right-hand panel shows violin plot for the distribution of the differences (Delta) of the four groups between the $\nu_{\rm d}$ of our measurements and the $\nu_{\rm d}$ from previous studies.
}
\label{fig:compare}
\end{figure*}

Many previous studies on the scintillation properties of pulsars have been published. The sources in our analysis overlap most with those of \citet{lmj16} and \citet{tmc+21}. Also, our data span almost matches their data span. It is, therefore, worthwhile to compare our scintillation parameters with theirs.
However, their scintillation parameters were determined at different radio frequencies. To compare them with our values, their $\nu_{\rm d}$ values were rescaled to our observing frequency by using a scaling index $\alpha$ = 4.4. For the scintillation timescale, we assumed that $\tau_{\rm d}$ is proportional to $f^{1.2}$ \citep{cl02}.
The comparison values are listed in Table~\ref{tab:scincompare}. 
The sub and super scripts are calculated by first adding and subtracting the uncertainties that were given in literature and then rescaling.

For a visual comparison, we plotted the $\nu_{\rm d}$ from previous works and our measurements from Table~\ref{tab:scincompare} in the left panel of Figure~\ref{fig:compare}. 
Both vertical and horizontal error bars of each point are asymmetric, the smallest and biggest values of error bars are the subscript and superscript in Table~\ref{tab:scincompare}.
Since we used the 5 and 95 percentiles of all measurements as sub and super scripts for each data set in this work in Table~\ref{tab:scincompare}, a whole horizontal error bar indicates the range of 90\% of the probability density. The gray line is the equivalence diagonal, which means that those points where our measurements are in keeping with the previous works lie on or near this line. 
Comparing data at both of our center frequencies with both literature sources, we created four groups and plotted them with a different color. 

To examine the probability density of the difference between our measurements and those from literature for the four groups we mentioned in previous paragraph, we plotted four violin plots of the differences, Delta, in the right panel of Figure~\ref{fig:compare}, where
\begin{equation}
\label{eq:delta}
{\rm Delta} = \frac{M_{\rm this~work}-M_{\rm literature}}{\sqrt{\sigma_{\rm this~work}^2 + \sigma_{\rm literature}^2}}
\end{equation}
with $M$ the median of the $\nu_{\rm d}$ for a given pulsar and at a given observing frequency, and $\sigma$ is its measurement uncertainty. We used the standard deviation of our measurements as the uncertainty for our own measurements. For the values from literature we determined the upper (lower) uncertainty as the difference between the superscripts (respectively the subscripts) and the most likely value quoted in Table~\ref{tab:scincompare}. These upper and lower uncertainties are then combined by taking the root-mean-square (rms) of them. This rms value is used as $\sigma_{\rm literature}$ in Equation~\ref{eq:delta}.

Comparison of the scintillation bandwidths in both panels of Figure~\ref{fig:compare} shows that, at the 21-cm band, our values are typically consistent with those from literature, especially for \citet{tmc+21}. In the 11-cm band, the previously published values show a notable bias towards small values compared to the values from our work, but they are still consistent within their error bars. 
There are some possible reasons to explain those discrepancies: 
\begin{enumerate}
    \item The strength of scintillation changes over time, as the time series of scintillation bandwidth of PSR~J0636+5128 in Figure~\ref{fig:TS} illustrates, where the values after MJD\,57600 increased significantly.
    \item Scaling indices can be expected to be different for different pulsars, and may also vary with time \citep{bts+19}. A scaling index of $\alpha$ = 4.4 based on the classical Kolmogorov spectrum represents a mean value for all pulsars, but most pulsar lines of sight do not follow a pure Kolmogorov spectrum and show a smaller or larger value of scaling index. So errors can easily be introduced when we rescale the scintillation bandwidth to another frequency using a scaling index of $\alpha$ = 4.4; the size of these errors is proportional to the frequency range scaled. That is why there is a larger bias at the 11-cm band.
    \item Unreliable measurements affected by limited frequency resolution and limited bandwidth also contribute to discrepancies. Usually, a higher frequency resolution and a wider bandwidth will produce more accurate measurements.
    \item The different data analysis methods can also introduce slight differences.
\end{enumerate}

Usually, the fourth reason only leads to a tiny discrepancy that is within the error bar. For the different data span, the first reason may lead to a significant discrepancy. For example, PSR~J0636+5128 show a visible increase round MJD\,57600 in the time series of $\nu_{\rm d}$ and $\tau_{\rm d}$, and the previously published works did not cover data after 2017. In this case, the increased monitoring length introduces a discrepancy. Luckily, our data span almost fully covers the data span of \citet{lmj16} and \citet{tmc+21}, and does not extend much beyond theirs, so that in the majority of cases this first reason does not cause discrepancies.

However, there are some significant discrepancies between our work and previous publications, which are highlighted by three ellipses in the left panel of Figure~\ref{fig:compare}.
The points in the red ellipse show that the previous measurements are larger than our own, which is caused by the third reason, the limited frequency resolution. Since we have a higher resolution for the JBO and WSRT, we can get a more accurate measurement when scintles are very narrow in frequency. The points in the blue ellipse are more likely caused by the second reason that is an incorrect scaling index. For the points in the yellow ellipse, the second and third reasons both are responsible for the significant discrepancies.

The finite observation length is the main cause restricting the accuracy of $\tau_{\rm d}$ in both our own and previously published analyzes.
All observations of \citet{lmj16,tmc+21} were from NANOGrav and were $\sim$30 minutes in duration, which makes the calculation of $\tau_{\rm d}$ problematic. Consequently, \citet{lmj16} did not present values for $\tau_{\rm d}$, while \citet{tmc+21} only gave $\tau_{\rm d}$ for a few pulsars. Most of our observations have a duration of longer than 30 minutes, which allows us to get reliable $\tau_{\rm d}$ values for more pulsars.

There are four pulsars that have $\tau_{\rm d}$ values derived in both \citet{tmc+21} and our work. Comparison of these values displays slight discrepancies, which are caused by a number of reasons (identical or similar to those mentioned for $\nu_{\rm d}$): 
(a) the strength of scintillation changes over time, (b) an incorrect scaling index, (c) the limited observation duration, (d) the different data analysis methods used. In addition, $\tau_{\rm d}$ is dependent on the Earth's velocity and pulsar orbital velocity, both of which vary with time. Therefore, for the limited number of observations, the different observing times also likely cause some discrepancy.

We also tried using another widely used scaling, 4.0 for the $\nu_{\rm d}$ and 1.0 for the $\tau_{\rm d}$, to rescale the previous publish values. Although there are some tiny offsets in the rescale values, the conclusion about the comparison has not been changed.

\subsection{Comparison with NE2001}
\begin{figure}
\centering
\includegraphics[width=10cm]{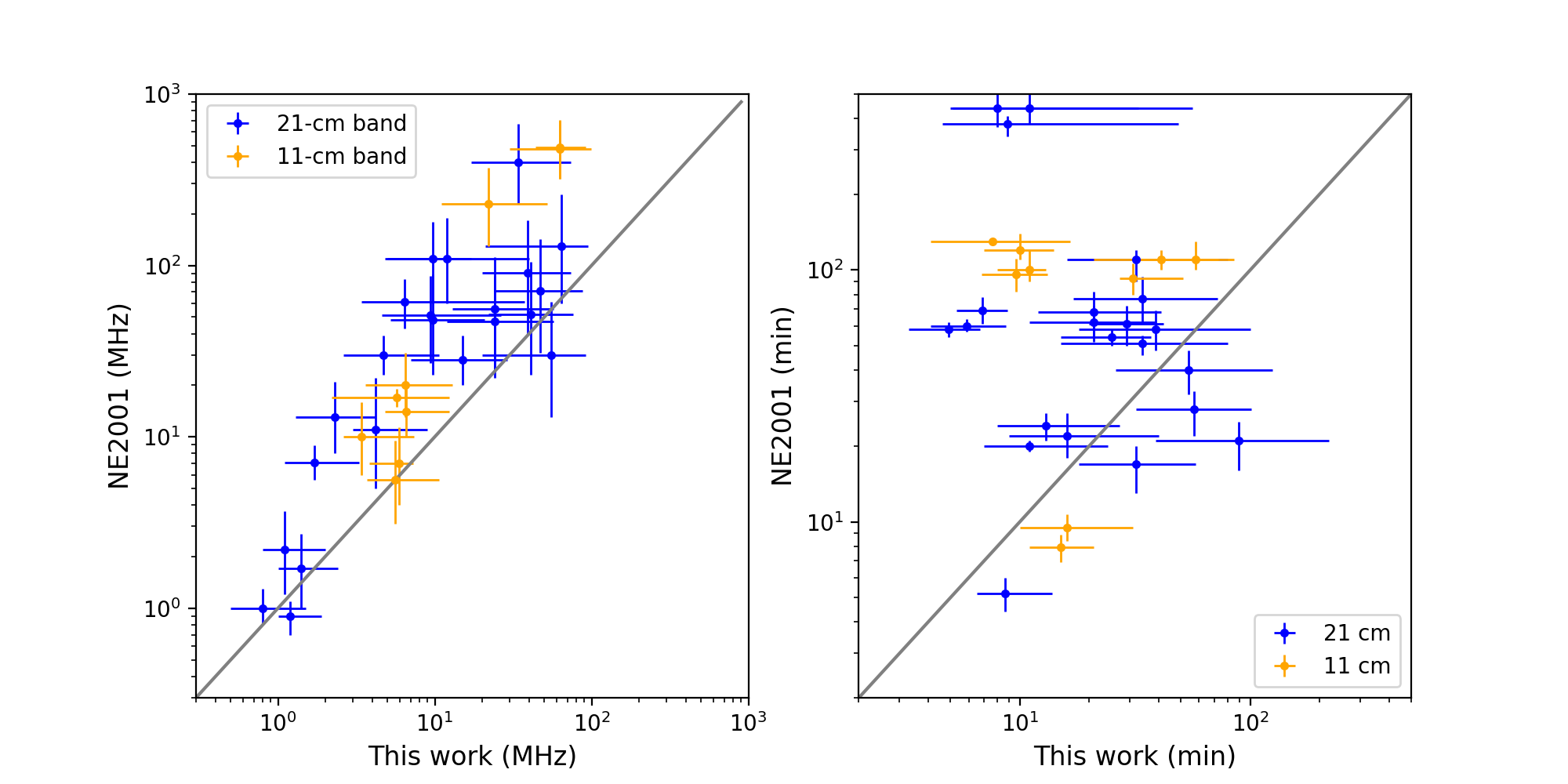}
\caption{The NE2001 predictions of scintillation parameters vs our measurements. The left panel shows the comparison of $\nu_{\rm d}$, the right panel shows the comparison of $\tau_{\rm d}$. 
}
\label{fig:comparene2001}
\end{figure}

We also compared our measurements with the predicted scintillation parameters from NE2001 \citep{cl02} which is a model for the Galactic distribution of free electrons and can be used to estimate the scintillation properties for a given sky position and DM value. Since NE2001 only gives the predictions at 1 GHz, we used a scaling index of $\alpha$ = 4.4 to rescale $\nu_{\rm d}$ to our observing frequencies. Note that the value of $\tau_{\rm d}$ depends not only on the frequency but also on the pulsar transverse velocity, as deduced in Equation (46) of \citet{cl91}. 
In the prediction of $\tau_{\rm d}$, the NE2001 Electron Density Model assumes that all pulsars have a pulsar transverse velocity of 100 km\,s$^{-1}$. In practice the pulsar transverse velocity ranges from a few km\,s$^{-1}$ to several thousand km\,s$^{-1}$. 
In addition to using a frequency scaling index of 1.2, the estimated pulsar transverse velocity also has been used to rescale $\tau_{\rm d}$, as in Equation (46) of \citet{cl91}.
Here the pulsar transverse velocity was calculated by
\begin{equation}
\label{eq:vp} 
V_{\rm \mu} = 4.74 \mu D_{\rm{kpc}},
\end{equation}
where $V_{\rm \mu}$ is in units of km\,s$^{-1}$, $\mu$ is the proper motion with units of mas\,yr$^{-1}$, and $D_{\rm{kpc}}$ is the pulsar distance in kpc. All the reference values and reference papers are listed in Table~\ref{tab:velocity}. 

\begin{table}
  \centering 
  \caption{Pulsar transverse velocity data and references.}
  \label{tab:velocity}
  \begin{tabular}{cccccc}
  \hline
  \hline
  Pulsar & $\mu$ & D & reference & $V_{\rm{pm}}$\\
         &  mas\,yr$^{-1}$& kpc & ($\mu$, D) & km\,s$^{-1}$\\
\hline
J0023+0923 & 13.87 & 1.10 & (1, 1) & 72.32 \\
J0613$-$0200 & 10.51 & 1.10 & (3, 2) & 54.82 \\
J0636+5128 & 4.74 & 0.20 & (4, 4) & 4.56 \\
J1022+1001 & 15.93 & 0.72 & (5, 5) & 54.37 \\
J1600$-$3053 & 7.06 & 3.00 & (2, 2) & 100.45 \\
J1640+2224 & 11.49 & 1.52 & (9, 9) & 82.79 \\
J1713+0747 & 6.29 & 1.05 & (3, 6) & 31.29 \\
J1857+0943 & 6.06 & 1.20 & (6, 7) & 34.50 \\
J1939+2134 & 0.41 & 5.00 & (3, 8) & 9.65 \\
J2145$-$0750 & 13.16 & 0.62 & (5, 5) & 38.67 \\
J2214+3000 & 20.64 & 0.40 & (1, 1) & 39.14 \\
J2234+0944 & 32.74 & 0.80 & (1, 1) & 124.13 \\
J2317+1439 & 4.08 & 1.66 & (5, 5) & 32.12 \\
\hline
\hline
  \end{tabular}
  \tablefoot{(1) \citet{abb+18}, (2) \citet{mnf+16}, (3) \citet{dcl+16}, (4) \citet{slr+14}, (5) \citet{dgb+19}, (6) \citet{vbc+09}, (7) \citet{vlh+16}, (8) \citet{vwc+12}, (9) \citet{vdk+18}}
\end{table}

Then the rescaled values of $\nu_{\rm d}$ and $\tau_{\rm d}$ from the predictions of NE2001 are listed in Table~\ref{tab:scincompare}, the subscripts and superscripts are the predicted values at the bottom and top of our bandwidth, respectively. Similar to the treatment of the comparison with previous works, we plotted the predictions from NE2001 vs our measurements in Figure~\ref{fig:comparene2001} to compare. According to the effective center frequency of observations, we created two groups: 11-cm band and 21-cm band.

As shown in Table~\ref{tab:scincompare} and Figure~\ref{fig:comparene2001}, our measurements for $\nu_{\rm d}$ are systematically smaller than the predictions for most pulsars. 
An incorrect scaling index or limits to our observing bandwidth and resolution could affect these results, although they cannot fully explain this apparent bias. An alternative explanation is a potential sample bias in our work, since the sample was defined by high NE2001 predictions for $\nu_{\rm d}$. Finally, it is possible that NE2001 overestimates scintillation bandwidth for nearby pulsars, although this cannot be rigorously determined based on our limited sample. 

For $\tau_{\rm d}$, some pulsars show that our measurements are smaller than predicted, some pulsars show our measurements are consistent with predictions, some pulsars show our measurements are larger than predicted. Except for an incorrect scaling index and the limited duration of our observations, an incorrect estimate of the pulsar transverse velocity can also contribute to significant discrepancies. In addition, almost all pulsars show that the range of variation from our measurements is larger than those from predictions, which is because, in the prediction of $\tau_{\rm d}$ of the NE2001 model, the Earth's velocity and pulsar orbital velocity have not been taken into consideration.


\section{Discussion}

\label{sec:discussion}

\begin{figure*}
\centering
\includegraphics[width=18cm]{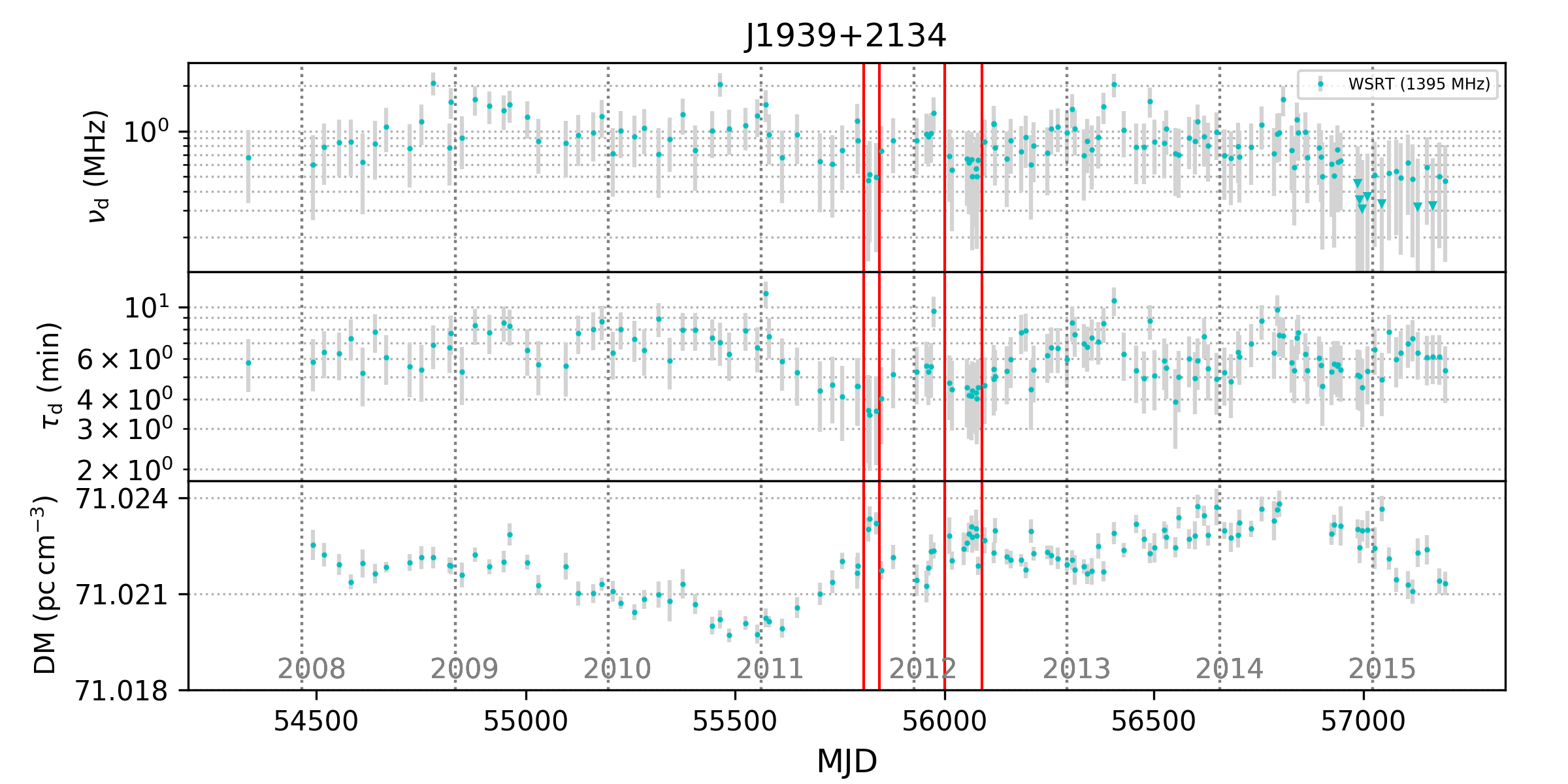}
      \caption{Time series of scintillation bandwidth $\nu_{\rm d}$, scintillation timescale $\tau_{\rm d}$ and DM for PSR~J1939+2134. There are two events, that show decreases in $\nu_{\rm d}$ and $\tau_{\rm d}$, and increases in DM, which are indicated by the red lines. The first one is between MJDs 55807 and 55845, the second one is between MJDs 56000 and 56090.}
      \label{fig:ese}
   \end{figure*}

Based on the measurements presented in Section~\ref{sec:results}, 
we found some interesting features in the time series of scintillation parameters. In this section, we describe future lines of research based on these features.

\subsection{Annual variations of the $\tau_{\rm d}$}

\begin{table*}
  \centering 
  \caption{The annual variation of the $\tau_{\rm d}$}
  \label{tab:annual}
  \begin{tabular}{ccc ccc ccc}
  \hline
  \hline
  Pulsar & DM           & ${\rm P_{b}}$ & A$_{1}$ & $V_{\rm o}$ & $V_{\rm \mu}$  & Annual variation \\
         & pc cm$^{-3}$ & days    & s       &km\,s$^{-1}$ & km\,s$^{-1}$  & \\
\hline
J0023+0923 & 14.3 &0.1388& 0.0348 & 5.5  & 72.3& No \\
J0613$-$0200 & 38.8 &1.1985& 1.0914 & 19.9  & 54.8& Strong  \\
J0636+5128 & 11.1 &0.0666& 0.009 & 2.9  & 4.6& Strong \\
J1022+1001 & 10.3 &7.8051& 16.7654 & 46.8  & 54.4& No  \\
J1600$-$3053 & 52.3 &14.3485& 8.8017 & 13.4  & 100.5& No \\
J1640+2224 & 18.4 &175.4607& 55.3297 & 6.9  & 82.8& No \\
J1713+0747 & 16.0 &67.8251& 32.3424 & 10.4  & 31.3& Weak \\
J1857+0943 & 13.3 &12.3272& 9.2308 & 16.3  & 34.5& No \\
J1939+2134 & 71.0 &$\ast$& $\ast$ & $\ast$  & 9.6& Weak \\
J2145$-$0750 & 9.0 &6.8389& 10.1641 & 32.4  & 38.7& No \\
J2214+3000 & 22.6 &0.4166& 0.0591 & 3.1  & 39.1& No \\
J2234+0944 & 17.8 &0.4197& 0.0684 & 3.6  & 124.2& No \\
J2317+1439 & 21.9 &2.4593& 2.3139 & 20.5  & 32.1& No \\
\hline
\hline
  \end{tabular}
  \tablefoot{The orbital period ${\rm P_{b}}$ and the projected semi-major axis ${\rm A_{1}}$ come from the ATNF Pulsar Catalogue\footnote{https://www.atnf.csiro.au/people/pulsar/psrcat/} \citep{mhth05}.}
\end{table*}

The time series of $\tau_{\rm d}$ shows a strong annual variation for PSRs~J0613$-$0200 and J0636+5128.
The scintillation timescale $\tau_{\rm d}$ is given by the ratio of the spatial scale of the diffractive scintillation pattern $l_{\rm d}$ and the scintillation velocity $V_{\rm{ISS}}$ \citep{cr98}:
\begin{equation}
\label{eq:taud}
\tau_{\rm d} = \frac{l_{\rm d}}{V_{\rm ISS}}.
\end{equation}
In a thin screen scattering geometry model, the scintillation velocity $V_{\rm ISS}$ can be determined by the effective transverse velocity $V_{\rm eff}$ and the fractional distance of the scattering screen $s$ (the pulsar at $s$ = 0 and the Earth at $s$ = 1), $V_{\rm ISS}$ = $V_{\rm eff}/s$ \citep{cr98}. The $V_{\rm eff}$ is a combination of the pulsar, Earth, and IISM velocities at scintillation screen position $s$,
\begin{equation}
\label{eq:veff}
V_{\rm eff}(s) = sV_{\rm E} + (1 - s)(V_{\rm p} + V_{\rm \mu}) - V_{\rm IISM}(s),
\end{equation}
where $V_{\rm E}$ is the Earth velocity, $V_{\rm p}$ is pulsar orbital transverse velocity and $V_{\rm IISM}$ is the IISM velocity. Equations~\ref{eq:taud} and \ref{eq:veff} clarify that variations in  $\tau_{\rm d}$ are determined by six parameters: $l_{\rm d}$, $s$, $V_{\rm E}$,$V_{\rm p}$, $V_{\rm \mu}$ and $V_{\rm IISM}$. Annual variations in $\tau_{\rm d}$ arise from the changes in the transverse velocity of the Earth.

Although the Earth's motion has contributions to $\tau_{\rm d}$ for all pulsars, not all pulsars showed annual variations in $\tau_{\rm d}$. 
Table~\ref{tab:annual} summarizes which pulsars show signs of annual variations in $\tau_{\rm d}$.  
To distinguish under which conditions annual variations can be observed, in Table~\ref{tab:annual} we also list $V_{\rm \mu}$ (which is the same as in Table~\ref{tab:velocity}) and the mean orbital velocity $V_{\rm o}$ (which was calculated based on the orbital period $P_{\rm b}$ and the projected semi-major axis ${\rm A_{1}}$, which in turn are available from pulsar timing). 
The orientation $\Omega$ and the inclination angle $i$ of the pulsar orbit are difficult to estimate and are known only for a few pulsars, therefore, we calculated the $V_{\rm o}$, not $V_{\rm p}$ in Table~\ref{tab:annual}.
From Table 4, one sees that all pulsars that showed annual variations had relatively small $V_{\rm \mu}$ and $V_{\rm o}$. But the reverse is not true: some pulsars with small $V_{\rm \mu}$ and $V_{\rm o}$ did not show annual variations, likely because of the location of the scattering screen $s$
reduces the amplitude of annual variations.
Indeed, from Equations~\ref{eq:taud} and \ref{eq:veff}, we can determine that  pulsars with a small $V_{\rm \mu}$, a small $V_{\rm p}$ and a large $s$ are more likely to show annual variations in $\tau_{\rm d}$.
In case of a large $s$, the contribution of pulsar velocities on $V_{\rm eff}$ has been reduced. On the other hand, a large $s$ indicates that the scintillation screen is close to the Earth. Plasma clouds close to the Earth are likely related to two features: 
the edge of the Local Bubble (LB)\footnote{The Local Bubble is a volume with a radius of around 100 pc, devoid of dense interstellar material and is roughly centered on the Sun \citep{lwv+03}.} and the local interstellar medium (LISM)\footnote{Within the Local Bubble, some warm plasma clouds are found in the immediate vicinity of the Sun and are often referred to as the LISM.}, and their velocities are supposed to be of the order of 10\,km\,s$^{-1}$.

PSRs~J0613$-$0200 and J0636+5128 both show a very strong annual variation in $\tau_{\rm d}$, which can be used to estimate the small-scale distribution along the line of sight and inhomogeneities of the IISM, like has been done by \citet{rcn+14} and \citet{rch+19}. This analysis will be published in a forthcoming paper.

\subsection{Correlations between $\nu_{\rm d}$ and $\tau_{\rm d}$}
The time series of scintillation parameters $\nu_{\rm d}$ and $\tau_{\rm d}$ show correlated variations in a number of cases:
\begin{enumerate}
 \item PSR~J0636+5128 with a strong annual variation in $\tau_{\rm d}$ shows also a weak annual variation in $\nu_{\rm d}$.
  \item The $\nu_{\rm d}$ and the $\tau_{\rm d}$ of PSR~J0636+5128 both show a visible increase along with periods of persistent highs around MJD~57600.
  \item The $\nu_{\rm d}$ and the $\tau_{\rm d}$ of PSR~J1939+2134 both show dips in the two regions between the vertical red lines of Figure~\ref{fig:ese}. These two events are most clearly detected in the WSRT data due to their high frequency resolution, so the WRST data are plotted separately in Figure~\ref{fig:ese}.
\end{enumerate}
The occurrence of correlated variations is not unexpected, since effectively both $\nu_{\rm d}$ and $\tau_{\rm d}$ depend on the scale of the diffractive scintillation pattern, $l_{\rm d}$, as described by \citet{cr98}. The consequence is, that if anisotropic scattering causes spatial variations in $l_{\rm d}$; and if either the screen is sufficiently close to Earth or the variability of $l_{\rm d}$ exists on a sufficiently small scale that the line of sight probes parts of the IISM with significantly different $l_{\rm d}$ over the course of a year, then annual variations would be expected in both $\tau_{\rm d}$ and $\nu_{\rm d}$. 

The dependence of $\tau_{\rm d}$ on $l_{\rm d}$ was already given in Equation~\ref{eq:taud}, the dependence of $\nu_{\rm d}$ on $l_{\rm d}$ is given by \citep{cr98}:
\begin{equation}
   \label{eq:ld}
\nu_{\rm d} \propto l^{2}_{\rm d}f^{2},
\end{equation}
where $f$ is the observing frequency.
Thus, if the annual variations enter $l_{\rm d}$, we can also expect an annual variation in the time series of $\nu_{\rm d}$. But the annual variation of $\nu_{\rm d}$ is far weaker than that of $\tau_{\rm d}$ because annual variations in $l_{\rm d}$ are likely small, while those in V$_{\rm ISS}$ can be large.

Equations~\ref{eq:taud} and \ref{eq:ld} also can be used to explain the second and third synchronous variations between $\nu_{\rm d}$ and $\tau_{\rm d}$. When some unpredictable changes happen for $l_{\rm d}$, which could happen when particular plasma clouds dominate the ISS at different times, $\nu_{\rm d}$ and $\tau_{\rm d}$ would both be affected and show similar changes in a similar time.

\subsection{Frequency dependence}
\begin{figure}
\centering
\includegraphics[width=9cm]{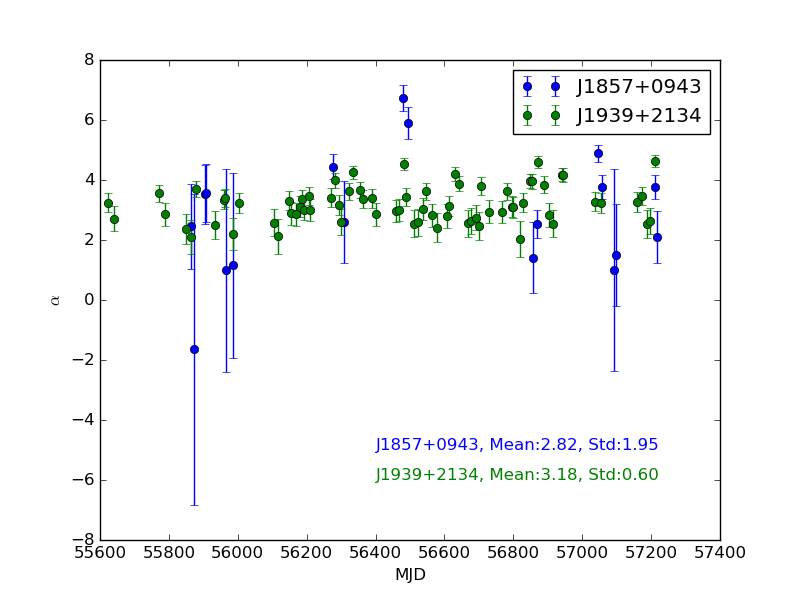}
\caption{The time series of frequency scaling indexes of the $\nu_{\rm d}$ for two pulsars. The mean and standard deviation of scaling indexes are printed in the right-bottom corner.}
\label{fig:scalingindex}
\end{figure}

After ruling out all limits of $\nu_{\rm d}$ and $\tau_{\rm d}$ due to the limited frequency resolution and the limited observing duration, 
as we expected, Table~\ref{tab:scinpara} and Figure~\ref{fig:TS} both show that measurements of $\nu_{\rm d}$ and $\tau_{\rm d}$ increase with observing frequency in general, and the frequency dependence of $\nu_{\rm d}$ is far stronger than that of $\tau_{\rm d}$. 
Since the frequency dependence of $\tau_{\rm d}$ is very weak and $\tau_{\rm d}$ is modulated by the Earth's and the pulsar's velocities which vary strongly with time, we only investigate the scaling with frequency of $\nu_{\rm d}$.
The strength of scintillation fluctuates with time and we do not have synchronous observations at the different observing frequencies. To remedy 
the impact of that, only measurements of $\nu_{\rm d}$ at the different bands and with a time interval to the closest observation of less than 7 days have been used to estimate the scaling index. Then, the frequency scaling index $\alpha$ of $\nu_{\rm d}$ is given by:
\begin{equation}
\label{eqn:scalefunc}
    \alpha = \frac{\log\left(\sfrac{\nu_{\rm d1}}{\nu_{\rm d2}}\right)}{\log\left(\sfrac{f_1}{f_2}\right)}
\end{equation}
where the $\nu_{\rm d1}$, $\nu_{\rm d2}$ and the $f_{1}$, $f_{2}$ are scintillation bandwidths and observing frequencies for two bands, respectively. Note that if a set at a given center frequency and a given telescope have 50\% or more limits (as opposed to measurements), we disregard the whole data set here.
In Figure~\ref{fig:scalingindex}, we plot the time series of the scaling index for PSRs~J1857+0943 and J1939+2134. One sees significant correlated variations, suggesting these variations are not random on these timescales. The mean and standard deviation of the $\alpha$ for PSRs~J1857+0943 and J1939+2134 are 2.82\,$\pm$\,1.98 and 3.18\,$\pm$\,0.60, respectively.

\subsection{The correlation between scintillation and DM}
\label{sec:corr_scin_dm}
In Figure~\ref{fig:TS}, we see two interesting instances of correlated variations between scintillation parameters and DM. Firstly, 
during the period of the two dips in $\nu_{\rm d}$ that are mentioned above in Figure~\ref{fig:ese}, the DM shows two increases. \citet{cks+15} detected similar events in PSRs~J1603$-$7202 and J1017$-$7156, they exploited those events to make more complete models of the extreme scattering event (ESE), including an estimate of the “outer-scale” of the turbulence in the plasma lens.
Secondly, similar to the two events mentioned above, in Figure~\ref{fig:ese}, PSR~J1939+2134 show an inverse relation between $\tau_{\rm d}$ and the DM through much of our data set, when $\tau_{\rm d}$ increases, the DM decreases, and vice versa.

In the lens model of scintillations \citep{pk+12,pl14}, changes in DM and the refracting angle $\delta$ of scintillation can both be used to infer the required electron column densities. \citet{lll+21} have already used this to calculate transient pulse profile changes expected to occur in concert with DM changes, and compared these with what is observed during two 'dip' events of DM for PSR~J1713+0747.
More details will be analyze in a forthcoming paper.

\subsection{Impacts on timing}
PTAs aim to detect gravitational waves through spatially correlated timing residuals between nearby millisecond pulsars \citep[see][and references therein]{vob+21}.
To successfully detect gravitational waves, corrupting effects have to be mitigated down to unprecedented levels. Interstellar scintillation is one effect that has been shown to have the potential to corrupt pulsar-timing experiments at levels significant to PTA science \citep{lkd+17}. 
Here we briefly interpret how scintillation impacts pulsar timing.

When pulsar signals travel through the inhomogeneous and highly turbulent IISM, another scintillation result, the scattering delay $\tau_{\rm st}$, can be calculated from 
\begin{equation}
\label{eq:C1}
2\pi\nu_{\rm d} \tau_{\rm st}= C_{1}
\end{equation}
\noindent where $C_{1}$ is a constant and close to unity (its value changes slightly for different geometries and models of the turbulence wavenumber spectrum, for a Kolmogorov spectrum, $C_1$ = 1.16) \citep{lr99}.
The $\tau_{\rm st}$ provided two potential ways to affect the timing precision \citep{crg+10}. Firstly, the pulse shape can be affected by pulse broadening: angular scattering broadens sharp single pulses into quasi-exponential pulses with a scattering delay $\tau_{\rm st}$, but how this affects the ToAs depends on the overall shape and complexity of the pulse profile. In this case, since fundamentally no distinction can be made between a time-constant amount of interstellar scattering and the intrinsic profile shape, the basic value for $\tau_{\rm st}$ is of no direct relevance to timing experiments. Instead, variations of $\tau_{\rm st}$ determine the impact on timing. Secondly, \citet{hs+08,msa+20} supposed that the shape of the pulse is effectively unchanged because the $\tau_{\rm st}$ is much less than the width of the pulse profile and pointed out that the leading order effect from scintillation would be a time-shift of the pulse profile.

The scattering delays $\tau_{\rm st}$ of 13 pulsars are calculated exploiting the measurements of $\nu_{\rm d}$ and the Equation~\ref{eq:C1}, we presented them in the last column of Table~\ref{tab:scinpara} by the median value and their 5/95 percentiles. 
All values range from a few nanoseconds to hundreds of nanoseconds, implying that for most pulsars in our sample the median amount of scattering is hard or impossible to determine at these frequencies. 
From just the lower and upper values of inferred scattering time, one
infers that the largest differences are of order 100 ns. This implies that for PTA efforts, scattering variations may be irrelevant in most pulsars, but could be relevant in some.

In order to study potential improvements to the timing precision, however, it would be more meaningful to determine the time variable impact on timing based on the observed variations in scintillation; and while taking into account the actual shape of the pulse profile. 
This can readily be quantified though simulations, as will be presented in a follow-up paper.

\section{Conclusions and further research}
\label{sec:conclusion}
In this paper, we presented measurements of $\nu_{\rm d}$ and $\tau_{\rm d}$ for 13 EPTA pulsars at multiple epochs by creating dynamic spectra and their ACFs. We plotted the time series of scintillation parameters and DM. We compared our scintillation parameters with previously published values and predictions from the NE2001 electron density model. The comparison showed that our results are typically consistent with those from  literature. More specifically, for some pulsars we obtained more precise measurements in $\nu_{\rm d}$ because we have higher frequency resolution at WSRT and JBO; since most of our observations have a longer observing length than \citet{lmj16} and \citet{tmc+21}, we were able to determine $\tau_{\rm d}$ for more pulsars. Furthermore, we have a longer time span of data and a larger number of observations for our pulsars, thus we were able to identify more interesting features in the time series of our scintillation parameters, as pointed out in Section~\ref{sec:discussion}. This will be discussed more fully in subsequent papers. In the comparison with NE2001, we found that the $\nu_{\rm d}$ we achieved in this work are mostly smaller than those predicted by NE2001. For PSRs~J1857+0943 and J1939+2134, the multiband measurements of $\nu_{\rm d}$ allowed us to analyze the frequency scaling index. 

Our results could help to improve the timing precision of PTA millisecond pulsars \citep{wsv+22}, by converting the scintillation bandwidth into scattering timescales $\tau_{\rm{st}}$ and then subtracting these from TOAs. Furthermore, our work helps to determine the required observing length and frequency resolution for PTA observations to allow such schemes to work.

We will further investigate several useful and important applications of these data. Specifically, deeper analysis of this data set is being undertaken along the following lines:
\begin{enumerate}
  \item Annual variations in the time series of scintillation timescales. It is possible to determine the distance and orientation of IISM clumps, but it is dependent on the physics of scattering and the lens models applied and hence these measurements can help to test the various models. 
  \item The impact of variable scattering on pulsar timing. In order to go beyond simplified theoretical models and fold in the complexity of the pulse profiles, a simulation of the impact on timing from the variable scattering delays (as derived from the scintillation bandwidth) is being undertaken.
  \item Correlations between scintillation parameters and DM,  mentioned in Section~\ref{sec:corr_scin_dm}.
  Since the DM precision is not high enough to see trends in many pulsars of this work, we did not find many convincing correlations. But, a couple of these pulsars also have LOFAR DM time series \citep{dvt+20}, so a combined analysis could be meaningful. 
\end{enumerate}
Our analysis of these aspects with appear in subsequent papers in this series.

\section*{Acknowledgments}
We thank Bill Coles for useful discussions and suggestions, Julian Donner for allowing us access to the code that is used to obtain the time series of DM, Eleni Graikou for her contribution to the pulsar observing with EFF.

Part of this work is based on observations with the 100-m radio telescope of the Max-Planck-Institut f\"ur Radioastronomie (MPIfR) at Effelsberg in Germany. 
Pulsar research at the Jodrell Bank Centre for Astrophysics and the observations using the Lovell Telescope are supported by a consolidated grant from the STFC in the UK. 
The Nan\c{c}ay Radio Observatory is operated by the Paris Observatory, associated with the French Centre National de la Recherche Scientifique (CNRS). We acknowledge financial support from ``Programme National de Cosmologie et Galaxies'' (PNCG) of CNRS/INSU, France. 
The Westerbork Synthesis Radio Telescope is operated by the Netherlands Institute for Radio Astronomy (ASTRON) with support from the Netherlands Foundation for Scientific Research (NWO).

JPWV acknowledges support by the Deutsche Forschungsgemeinschaft (DFG) through the Heisenberg programme (Project No. 433075039). YL acknowledges support from the China Scholarship Council (No. 201808510133). J. W. McKee is a CITA Postdoctoral Fellow: This work was supported by Ontario Research Fund—research Excellence Program (ORF-RE) and the Natural Sciences and Engineering Research Council of Canada (NSERC) [funding reference CRD 523638-18].

\bibliographystyle{aa}
\bibliography{journals,psrrefs,modrefs,crossrefs}

\end{document}